\documentclass[aip,jcp,reprint]{revtex4-2}

\usepackage{amsmath,amsthm,latexsym,amssymb,amsfonts,epsfig}
\usepackage{bm}
\usepackage{dsfont}
\usepackage[left = 1.35cm, right = 1.35cm, top = 2.2cm, right = 2.2cm]{geometry}

\usepackage{graphicx}

\usepackage{color}

\definecolor{OliveGreen}{rgb}{0.1, 0.4, 0.1}
\definecolor{candyapplered}{rgb}{1.0, 0.03, 0.0}

\usepackage{setspace}
\emergencystretch=\maxdimen
\def\Xint#1{\mathchoice
   {\XXint\displaystyle\textstyle{#1}}%
   {\XXint\textstyle\scriptstyle{#1}}%
   {\XXint\scriptstyle\scriptscriptstyle{#1}}%
   {\XXint\scriptscriptstyle\scriptscriptstyle{#1}}%
   \!\int}
\def\XXint#1#2#3{{\setbox0=\hbox{$#1{#2#3}{\int}$}
     \vcenter{\hbox{$#2#3$}}\kern-.5\wd0}}

\def\dashint{\Xint-}

\allowdisplaybreaks

\usepackage[colorlinks=true,citecolor=OliveGreen,linkcolor=OliveGreen,urlcolor=OliveGreen]{hyperref}

\usepackage{mathtools}
\usepackage{mathrsfs}

\usepackage{type1ec} %

\begin{document}

\title{Hydrodynamic Flow Field and Frictional Resistance Coefficient of a Disk Rotating Steadily in a Compressible Fluid Layer With Odd Viscosity on a Rigid Substrate}%

\author{Abdallah Daddi-Moussa-Ider}%
\email{abdallah.daddi-moussa-ider@open.ac.uk}
\thanks{Author to whom correspondence should be addressed.}
\affiliation{School of Mathematics and Statistics, The Open University, Walton Hall, Milton Keynes MK7 6AA, United Kingdom}

\author{Yuto Hosaka}%
\affiliation{Max Planck Institute for Dynamics and Self-Organization (MPI-DS), Am Fa\ss berg 17, 37077 G\"{o}ttingen, Germany}
\date{\today}%

\author{Elsen Tjhung}
\affiliation{School of Mathematics and Statistics, The Open University, Walton Hall, Milton Keynes MK7 6AA, United Kingdom}

\author{Andrej Vilfan}%
\affiliation{Jo\v{z}ef Stefan Institute, 1000 Ljubljana, Slovenia}

\begin{abstract}
We analyze the hydrodynamics of a rotating disk in a two-dimensional compressible fluid layer with odd viscosity. Unlike conventional fluids, odd viscosity introduces a radial flow component that can be directed either inward or outward, depending on its sign. This phenomenon is expected to significantly impact the hydrodynamic interactions between two rotating disks, potentially causing effective attraction or repulsion depending on the sign of the odd viscosity and the direction of rotation. Furthermore, we calculate the rotational resistance coefficient and find that odd viscosity increases this coefficient, regardless of its sign.
\end{abstract}

\maketitle

\section{Introduction}

Chiral active fluids are systems composed of remarkable constituents that violate both parity and time-reversal symmetries. They appear in a variety of active matter systems, where active rotational motion takes place through continuous injection of energy~\cite{shankar2022topological}.
Examples span a wide range of scales, from aggregates of membrane-embedded rotors~\cite{oppenheimer2022hyperuniformity}, swimming algae~\cite{drescher2009dancing} and marine embryos~\cite{tan2022odd} to bacteria in circular motion~\cite{lauga2006swimming}.
Even at more macroscopic scales, such chiral systems can be formed by self-spinning rotors~\cite{yang2020} or colloidal particles~\cite{soni2019odd, mecke2024chiral}.

In classical fluids where symmetries are conserved, transport phenomena in a variety of physical systems follow a strong principle known as the Onsager reciprocal relations~\cite{onsager1931reciprocal}.
These relations impose symmetry on the transport coefficients that connect thermodynamic forces and flows.
In fluid dynamics, they are related to the Lorentz reciprocal theorem~\cite{masoud2019} that imposes a symmetric form of the resistance tensor of immersed objects~\cite{doi2013soft}.
Fluids with broken symmetries, however, can have unusual anti-symmetric components in the transport coefficients that are called \textit{odd viscosity}~\cite{avron1998, banerjee2017, hosaka2022nonequilibrium, fruchart2023odd, mecke2024emergent}.
Odd viscosity is dissipationless, but it can affect the dissipation in the fluid through its influence on the flow~\cite{khain2022, hosaka2023lorentz, everts2024dissipative}. 
To date, odd viscosity has been observed exclusively in 2D fluids of spinning particles, such as metallic colloids and colloidal rods in an external magnetic field~\cite{soni2019odd, mecke2023simultaneous}.
The mechanisms that lead to odd viscosity have been studied in numerical or computational studies of chiral active fluids~\cite{hargus2020time, han2021fluctuating, lou2022odd, caporusso2024phase, ding2024odd}.  The concept of odd viscosity has also been extended to other analogous notions such as systems with odd diffusivity and odd elasticity~\cite{hargus2021odd, kalz2022collisions, scheibner2020odd, fruchart2023odd}.

Although experimental protocols have been developed to measure odd viscosity using excess boundary stresses~\cite{soni2019odd}, weak compressibility~\cite{mecke2023simultaneous}, and the area change of a bubble~\cite{ganeshan2017} or even directly measuring the stress tensor of the fluid~\cite{hargus2020time, han2021fluctuating, zhao2021}, simple protocols, e.g., using a single probe particle, are scarce.
Therefore, methods to characterize chirality in odd fluids are not yet fully developed.
One of the potential approaches is to analyze the response of probes, a method referred to as microrheology~\cite{furst2017microrheology}.
In classical fluids, this method determines the transport properties of various types of soft media, such as monolayers of soluble surfactants~\cite{zell2014surface}, compressible fluids~\cite{barentin1999}, and viscoelastic environments~\cite{crocker2000two}.
Especially in compressible fluids with odd viscosity, rotational motion of particles has been less studied, as compared to translational motion~\cite{lier2023lift, duclut2024probe}.

Here, we study the hydrodynamics of a compressible fluid layer with odd viscosity surrounding a rotating disk.
We solve the problem analytically and determine the flow field around the disk.
Notably, the flow field contains a radial component, in addition to well-known azimuthal part, in stark contrast to the swirling flow in classical fluids.
Using this analytic solution, we also compute the rotational friction. This result offers a straightforward experimental setup for detecting odd viscosity, a key characteristic of chiral active fluids.

\section{2D Odd hydrodynamics}

\begin{figure}
    \centering
    \includegraphics[width=0.9\linewidth]{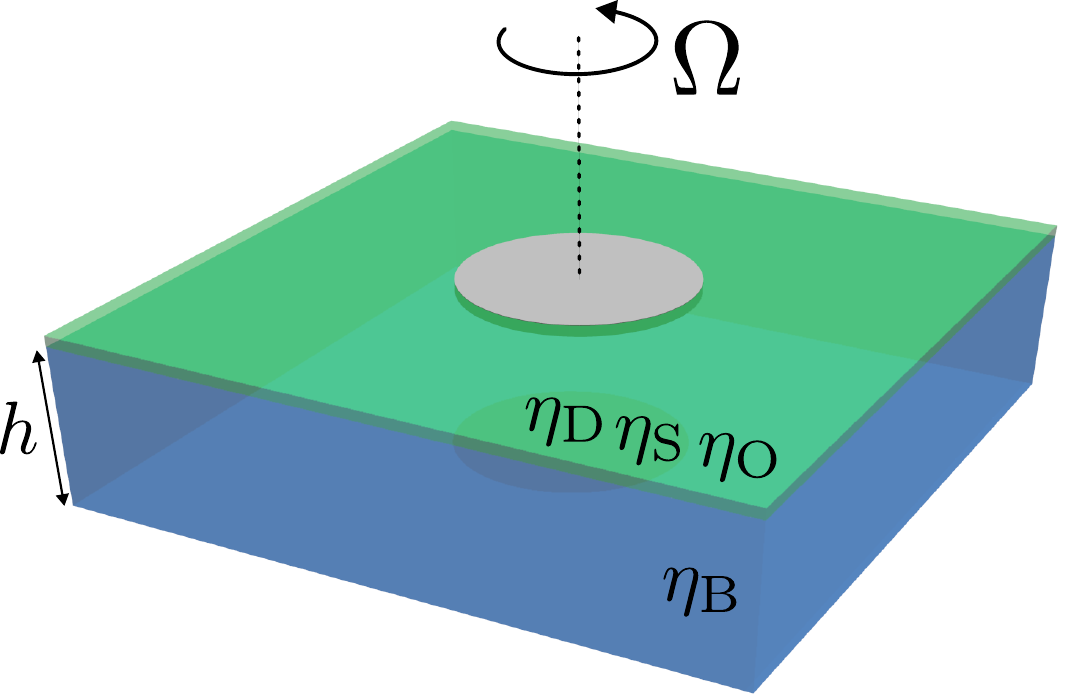} 
    \caption{
    (Color online) Schematic of a disk with radius $a$, rotating at a constant angular velocity $\Omega$, immersed in a thin fluid layer characterized by two dissipative viscosities (dilatational viscosity $\eta_{\rm D}$ and shear viscosity $\eta_{\rm S}$) and a non-dissipative odd viscosity $\eta_{\rm O}$. The 3D bulk fluid, which supports the overlying fluid layer from below, has only shear viscosity $\eta_{\rm B}$, with the subphase thickness denoted by $h$.
    In actual calculations, we use the lubrication approximation for the subphase and the thickness $h$ is much smaller than any in-plane length scales in the system, which imposes the condition $a\gg h$. Therefore, the comparable size of a disk and the thickness in the schematic is for presentation purposes only.
    }
    \label{fig:system}
\end{figure}

We consider a 2D compressible thin layer supported by an underlying incompressible 3D fluid; see Fig.~\ref{fig:system} for an illustration of the system setup. 
The 2D fluid layer is characterized by three viscosity coefficients, namely, dilatational~$\eta_{\rm D}$, shear~$\eta_{\rm S}$, and odd viscosity~$\eta_{\rm O}$, while the 3D bulk fluid possesses only a 3D shear viscosity~$\eta_{\rm B}$.
We assume that these viscosities are constant.
Let $\bm{f}_{\rm B}$ denote the force density exerted on the 2D  layer by the 3D fluid beneath it. Additionally, let $\bm{f}$ represent any other traction force acting on the 2D layer. 

The equation governing the in-plane fluid velocity $\bm{v}(x,y)$ on the thin layer is then given by the 2D compressible Stokes equation as
\begin{align}
    \boldsymbol{\nabla} \cdot \boldsymbol{\sigma} + \bm{f}_{\rm B} + \bm{f} = \mathbf{0} \, .
    \label{eq:balance}
\end{align}
Here, $\boldsymbol{\nabla} = (\partial_x, \partial_y)$ represents the 2D differential operator, and $\boldsymbol{\sigma}$ denotes the 2D viscous stress tensor
\begin{align}
\sigma_{ij} = \eta_{ijk\ell} E_{k\ell} \, ,
\end{align}
where $E_{ij} = \left( \partial_i v_j + \partial_j v_i \right)/2$ is the strain rate tensor and $\eta_{ijkl}$ is the viscosity tensor, which, in the case of a 2D compressible system~\cite{epstein2020}, can be represented as
\begin{align} \label{eq:viscosity}
    \eta_{ijk\ell} &= \left( \eta_{\rm D} - \eta_{\rm S} \right) \delta_{ij}\delta_{k\ell} 
    + \eta_{\rm S} (\delta_{ik}\delta_{j\ell}+\delta_{i\ell}\delta_{jk}) \notag  \\
    &\quad 
    +\frac{1}{2}\, \eta_{\rm O}(\epsilon_{ik}\delta_{j\ell}+\epsilon_{j\ell}\delta_{ik}+\epsilon_{i\ell}\delta_{jk}+\epsilon_{jk}\delta_{i\ell}) \, .
\end{align}
Here, $\delta_{ij}$ denotes the Kronecker delta and $\epsilon_{ij}$ stands for the 2D Levi-Civita tensor, which is antisymmetric with respect to $i \leftrightarrow j$, such that $\epsilon_{xx} = \epsilon_{yy} = 0$ and $\epsilon_{xy} = -\epsilon_{yx} = 1$.
As shown in Eq.~(\ref{eq:viscosity}), the odd viscosity term proportional to $\eta_{\rm O}$ breaks the symmetry under the exchange $ij\leftrightarrow k\ell$.
In general, additional odd viscosity and regular viscosity terms may appear in Eq.~(\ref{eq:viscosity}), coupling the stress tensor to the vorticity tensor~\cite{epstein2020, fruchart2023odd, markovich2019chiral}. 
However, for simplicity, we assume that the dynamics of angular momentum are decoupled from linear momentum, so that the anti-symmetric stress can be neglected~\cite{markovich2019chiral}.

The velocity of the 3D bulk fluid, $\bm{u}(x, y, z)$, satisfies the 3D incompressible Stokes equation with the no-slip boundary condition $\bm{u}(x, y, 0) = 0$ at $z = 0$ and the continuity condition $\bm{u}(x, y, h) = \bm{v}(x, y)$ at $z = h$ (see Fig.~\ref{fig:system}). By applying the lubrication approximation for the bulk fluid~\cite{barentin1999, elfring2016surface, manikantan2020surfactant}, the coupled 2D compressible equation~\eqref{eq:balance} and 3D incompressible Stokes equations with the boundary conditions at $z = 0$ and $z = h$ can be solved. This brings us to~\cite{hosaka2021nonreciprocal}
\begin{align}
\boldsymbol{\eta}
\cdot \boldsymbol{\nabla}^2 \bm{v}
+ \eta_\mathrm{D}\boldsymbol{\nabla} (\boldsymbol{\nabla}\cdot\bm{v}) 
-\frac{h}{2} \, \boldsymbol{\nabla} p 
-\frac{\eta_\mathrm{B}}{h} \, \bm{v}
+\bm{f} = \bm{0} \, ,
\label{eq:heq}
\end{align}
with $\boldsymbol{\eta} = \eta_{\rm S} \mathds{1} + \eta_{\rm O} \boldsymbol{\epsilon}$. Here, $h$ denotes the thickness of the bulk fluid, which is assumed to be much smaller than any planar dimension of the system, and $p$ represents the pressure in the bulk fluid, which is independent of $z$ in the lubrication approximation. Note that there is no 2D pressure in the governing equation because we have assumed that material transport at the interface equilibrates faster than the time scale of the disk motion~\cite{barentin1999}.
The lubrication approximation further provides the following relationship between the surface velocity and the bulk pressure~\cite{barentin1999, elfring2016surface}
\begin{equation}
    \boldsymbol{\nabla} \cdot\bm{v} = \frac{h^2}{6\eta_\mathrm{B}}\boldsymbol{\nabla}^2p \, .
    \label{eq:compressibility}
\end{equation}
Due to the coexistence of 2D and 3D viscosities in the system, the hydrodynamic screening lengths can be defined as
\begin{equation}
    \kappa^{-1} = \sqrt{\frac{h \eta_\mathrm{S}}{\eta_\mathrm{B}}} \, , \qquad 
    \lambda^{-1} = \sqrt{
    \frac{h\bar{\eta}}{2\eta_\mathrm{B}}} \, ,
\label{kappa_lambda}
\end{equation}
where we have defined the averaged even viscosity as $\bar{\eta} = \left(\eta_{\rm S} + \eta_{\rm D} \right) / 2$. 
For the validity of the lubrication approximation, additional conditions on these length scales—namely \( \kappa^{-1} \gg h \) and \( \lambda^{-1} \gg h \)—need to be imposed.~\cite{barentin1999}
Since momentum is transported beyond these length scales, the viscosity of the surface layer dominates over bulk when $r < \kappa^{-1}$ and $r < \lambda^{-1}$, while the effect of the supporting layer becomes dominant in the opposite limits.
For further details on the model and the assumptions made, we refer the reader to Refs.~\onlinecite{hosaka2021nonreciprocal,hosaka2023hydrodynamics,daddi2025analytical}.

For later reference, we define the dimensionless parameters
\begin{equation}
    \mu = \frac{\eta_\mathrm{O}}{\eta_\mathrm{S} } \, , 
    \qquad
    \xi = \frac{\kappa}{\lambda} =\sqrt{\frac{\bar\eta}{2 \eta_\mathrm{S}}}\, .
\end{equation}
In particular, $\mu$ measures how far the 2D layer deviates from classical fluids with conserved symmetries. Since the even viscosities are strictly positive due to the positivity of entropy production, the condition $\xi \geq 1/2$ holds, while the odd-to-even viscosity ratio, $\mu$, can be either positive or negative.

\section{Green's function}

The hydrodynamic response of the 2D fluid layer to external forces is described by the velocity Green's function \( \boldsymbol{\mathcal{G}}(\bm{r}) \), which links the applied force to the induced velocity:
\begin{align}
    \bm{v}(\bm{r})\
    =
    \int \boldsymbol{ \mathcal{G} }(\bm{r}-\bm{r}^\prime)\cdot\bm{f}(\bm{r}^\prime) \, \mathrm{d}^2 \bm{r}^\prime \, .
\end{align}
Similarly, we introduce the Green's function for the pressure
\begin{align}
    p(\bm{r})\
    =
    \int \boldsymbol{ \mathcal{P} }(\bm{r}-\bm{r}^\prime)\cdot\bm{f}(\bm{r}^\prime) \, \mathrm{d}^2 \bm{r}^\prime \, .
\end{align}

We employ a 2D Fourier transform approach to describe the evolution of hydrodynamic fields, a technique widely used to solve various flow problems in the low Reynolds number regime~\cite{daddi2016hydrodynamic, felderhof2006dynamics, daddi2018brownian, daddi2018hydrodynamic, daddi2019frequency}. The Green's function is obtained from the hydrodynamic equations~\eqref{eq:heq} and~\eqref{eq:compressibility} in Fourier space by expressing the fields in terms of the wavevector \( \bm{k} = (k_x, k_y) \). 
We introduce the two orthogonal unit vectors, \( \bm{k}_\parallel = \left(k_x, k_y\right)/k \) and \( \bm{k}_\perp = \left(-k_y, k_x\right)/k \), where \( k = |\bm{k}| \) denotes the wavenumber.
Using polar coordinates, where \( k_x = k \cos\phi \) and \( k_y = k \sin\phi \), we have \( \bm{k}_\parallel = \left( \cos\phi, \sin\phi \right) \) and \( \bm{k}_\perp = \left( -\sin\phi, \cos\phi \right) \).
By defining the scaled wavenumber \( u = k / \kappa \), we can express the Green's function for the velocity field as~\cite{hosaka2021nonreciprocal}
\begin{equation}
\hspace{-0.2cm}
 \widetilde{\boldsymbol{ \mathcal{G} }} (u,\phi) =
    \frac{5\left( \alpha u^2+1 \right) \mathds{1}
    -3\left( \beta u^2+1\right)
    \mathds{R}_\phi
    -2\mu u^2 \boldsymbol{\epsilon}}
    { 2\kappa^2 \eta_\mathrm{S} \big( 4 \left( u^2+1 \right) \left( \xi^2 u^2+1\right) + \mu^2 u^4 \big)} ,
\label{eq:G}
\end{equation}
where functions in the Fourier transform domain are denoted with a tilde.
We define the rotation matrix 
\begin{equation}
    \mathds{R}_\phi
    = \begin{pmatrix}
        \cos 2\phi & \sin2\phi \\[3pt]
        \sin2\phi & -\cos2\phi
    \end{pmatrix} ,
\end{equation}
together with the dimensionless numbers $\alpha = \left( 4\xi^2+1 \right)/5= \left( \eta_\mathrm{D}+2 \eta_\mathrm{S} \right)/ \left(5 \eta_\mathrm{S} \right)$ and $\beta = \left( 4\xi^2-1 \right)/3 = \eta_\mathrm{D}/\left( 3 \eta_\mathrm{S} \right)$, such that $\alpha \ge 2/5$ and $\beta \ge 0$.

Likewise, the Green's function for the pressure is derived from Eq.~\eqref{eq:compressibility} as
\begin{equation}
    \widetilde{\bm{\mathcal{P}}}(u,\phi) = 
    \frac{6i}{\kappa h u}
    \frac{ \mu u^2\, \bm{k}_\perp -(u^2+1) \, \bm{k}_\parallel }{4 \left( u^2+1 \right) \left( \xi^2 u^2+1\right) + \mu^2 u^4} \, .
    \label{eq:pressure-fourier}
\end{equation}

\section{Hydrodynamic fields}

We theoretically investigate the steady rotational motion of a circular disk rotating in a 2D fluid layer with odd viscosity, as shown in Fig.~\ref{fig:system}. 
The disk, with radius~\( a \), rotates with angular velocity~\( \Omega \). In polar coordinates, we define~\( r \) as the radial distance and \( \theta \) as the azimuthal angle. A no-slip boundary condition is imposed at the disk surface. 
We proceed to derive the resulting velocity and pressure fields around the disk by analytically solving the 2D odd Stokes equations \eqref{eq:heq} and~\eqref{eq:compressibility}.
Using the screening length, we define the scaled disk radius and polar distance as \( b = \kappa a \) and \( \rho = \kappa r \), respectively.

To determine the solution for the flow field induced by the steady rotation of the disk, we proceed as follows.
First, we propose the form of the force density, incorporating unknown coefficients~\cite{Daddi-Moussa-Ider_2024_JPCM, daddi2024rotational, daddi2024hydrodynamic_JFM}. Then we determine these coefficients in Fourier space using the velocity boundary condition at the surface of the rotating disk.

In cylindrical polar coordinates, the force density is expressed using the Ansatz
\begin{equation}
    \bm{f}(r,\theta) = \frac{\Omega \eta_\mathrm{S}  }{a} \left( f_\parallel(r) \, \hat{\bm{e}}_r + f_\perp(r) \, \hat{\bm{e}}_\theta \right) , \label{eq:force_real}
\end{equation}
where the radial and azimuthal components of the force density are assumed to have the forms
\begin{subequations}
    \begin{align}
    f_\parallel (r) &= A \, a \, \delta(r-a) \, , \\
    f_\perp (r) &= B \, a \,  \delta(r-a) 
    + C \, \frac{r}{a} \, \Theta(a-r) \, ,
\end{align}
\end{subequations}
with $\Theta$ representing the Heaviside function, and $A$, $B$, and $C$ being unknown coefficients to be determined later.
Accordingly, the force includes a Dirac contribution concentrated along the perimeter of the disk, along with a constant azimuthal component across its surface.

The Fourier transform of the force density, as given by Eq.~\eqref{eq:force_real}, is expressed as
\begin{equation}
    \widetilde{\bm{f}} (u, \phi) = 
    -2i\pi a\Omega \eta_\mathrm{S} \left( F_\parallel(u) \, \hat{\bm{k}}_\parallel
    + F_\perp(u) \, \hat{\bm{k}}_\perp
    \right) ,  \label{eq:force_fourier}
\end{equation}
where the longitudinal and transverse components in Fourier space, defined as parallel and perpendicular to the wavevector, respectively, are given by
\begin{subequations}
    \begin{align}
    F_\parallel(u) &= A J_1(bu) \, , \\
    F_\perp(u) &= B J_1(bu) + \frac{C}{bu} \, J_2(bu)\, ,  
\end{align}
\end{subequations}
with $J_\nu$ representing the $\nu$th order Bessel function of the first kind~\cite{abramowitz2000handbook}.

\subsection{Flow velocity field}

Substituting the Fourier-transformed force density from Eq.~\eqref{eq:force_fourier} into
\begin{equation}
    \widetilde{\bm{v}} = \widetilde{\boldsymbol{\mathcal{G}}} \cdot \widetilde{\bm{f}} \, ,
\end{equation}
with the Green's function \(\widetilde{\boldsymbol{\mathcal{G}}}\) given by Eq.~\eqref{eq:G}, and applying the inverse Fourier transform~\cite{baddour2011two}, the radial and azimuthal velocities can be cast in the integral form
\begin{equation}
    v_j(\rho) = a\Omega \int_0^\infty
    \frac{ \mathcal{K}_j (u) \, J_1(\rho u) \, \mathrm{d}u }{ 4(u^2+1)(\xi^2 u^2+1) + \mu^2 u^4 } \, , 
\end{equation}
with $j \in \{\rho, \theta\}$, where we have defined 
\begin{subequations}
    \begin{align}
    \mathcal{K}_\rho (u) &= u F_\parallel(u) + u^3 \left( F_\parallel(u) - \mu F_\perp(u) \right), \\[3pt]
    \mathcal{K}_\theta (u) &= 4u F_\perp(u)
    + u^3 \left( 4\xi^2 F_\perp(u) + \mu F_\parallel(u) \right) .
\end{align}
\end{subequations} 

The three unknown constants can be determined from the velocity boundary condition.
At the surface of the disk, the no-slip velocity condition is required, leading to
\begin{equation}
    \int_0^\infty
    \frac{ \mathcal{K}_j (u) \, J_1(\rho u) \, \mathrm{d}u }{ 4(u^2+1)(\xi^2 u^2+1) + \mu^2 u^4 } = 
\frac{\rho}{b} \, \delta_{\rho j} \, ,
\end{equation}
for $0 \le \rho \le b$ with $ j \in \{\rho,\theta\}$.

The determination of the constant requires the evaluation of nontrivial improper integrals. These integrals can, in fact, be evaluated exactly, yielding precise analytical expressions for the constants defining the force density.
We define the improper integrals 
\begin{equation}
    Z_n^q = \int_0^\infty \frac{u^q J_n(bu) J_1(\rho u) \, \mathrm{d} u}{4(u^2+1)(\xi^2 u^2+1) + \mu^2 u^4} \, ,
    \label{eq:Znq}
\end{equation}
for $(n,q) \in \{ (1,1), (1,3), (2,0), (2,2) \} \, .$

We define the column vectors \( \bm{\mathcal{V}} = \left( v_\rho, v_\theta \right)^\top / \left( a\Omega \right)\) and \( \bm{\mathcal{X}} = \left( A, B, C \right)^\top \).
The dimensionless velocity field solution can be expressed as
\begin{equation}
    \bm{\mathcal{V}}
    = \bm{\mathcal{A}} \cdot \bm{\mathcal{X}} \, ,
\end{equation}
with the $2\times 3$ matrix
\begin{equation}
    \bm{\mathcal{A}} = 
    \begin{pmatrix}
        Z_1^1+Z_1^3 & -\mu Z_1^3 & -\cfrac{\mu}{b} \, Z_2^2 \\[3pt]
        \mu Z_1^3 & 4 \left( Z_1^1 + \xi^2 Z_1^3 \right) & \cfrac{4}{b} \left( Z_2^0 + \xi^2 Z_2^2 \right)
    \end{pmatrix} . \label{eq:A_MATRIX}
\end{equation}

The integral defined by Eq.~\eqref{eq:Znq} is evidently convergent.
It can be evaluated analytically using the method of residues~\cite{spiegel2009complex}.

We define
\begin{equation}
    A_\pm = \left( \frac{2 \left( 1+\xi^2 \pm \varsigma^{\frac{1}{2}} \right)}{\mu^2 + 4 \xi^2} \right)^\frac{1}{2} ,
\end{equation}
where
\begin{equation}
    \varsigma = \left( 1-\xi^2\right)^2-\mu^2 \, .
\end{equation}
After some algebra, we find that the improper integrals \( Z_n^q \) can be calculated as
\begin{equation}
    Z_n^q = W_n^q(A_+) - W_n^q(A_-) + K_n^q \, , 
    \label{eq:Znq1}
\end{equation}
with 
\begin{equation}
\hspace{-0.2cm}
    W_n^q(z) = \frac{\pi (iz)^{q-1}}{8 \varsigma^\frac{1}{2}}
    \begin{cases}
         H_n^{(1)}(ibz) \, I_1(\rho z) \qquad \quad\text{if~} \rho \le b \\[3pt]
        i^{n-1} I_n(bz) \, H_1^{(1)}(i\rho z ) \, \quad \text{if~} \rho \ge b
    \end{cases} 
\end{equation}
and
\begin{equation}
    K_n^q = \frac{\rho}{4b^2} \, \Theta(b-\rho)  \, \delta_{n,2} \, \delta_{q,0} \, , \label{eq:Knq}
\end{equation}
where $H_n^{(1)}$ and $I_n$ denote the $n$th-order Hankel function and modified Bessel function of the first kind, respectively,~\cite{abramowitz2000handbook}.
Here, we have adopted the standard definition of the Heaviside function, with $\Theta(0^-) = 0$ and $\Theta(0^+) = 1$. Mathematical details outlining the derivation steps are provided in the Appendix.

\subsubsection{Solution inside the disk}

By inserting the expressions of~$Z_n^q$, determined by Eq.~\eqref{eq:Znq1}, into Eq.~\eqref{eq:A_MATRIX} for $\rho \le b$, corresponding to the domain inside the disk, we obtain
\begin{equation}
    \bm{\mathcal{A}} = \bm{\mathcal{M}}_+ \,I_1(\rho A_+) - \bm{\mathcal{M}}_- \,I_1(\rho A_-) + \bm{\mathcal{C}} \, ,
\end{equation}
where
\begin{equation}
     \bm{\mathcal{M}}_\pm =
     \frac{\pi}{8\varsigma^\frac{1}{2}}
     \left(
     \begin{matrix}
         ~\bm{\mathcal{M}}_1^\pm ~\large|~ \bm{\mathcal{M}}_2^\pm~
     \end{matrix}
     \right),
\end{equation}
with
\begin{subequations}
    \begin{align}
    \bm{\mathcal{M}}_1^\pm &= 
    \left(
       ~\bm{Z}_\pm ~ \large| ~ \bm{Q}_\pm~
    \right)
    H_1^{(1)}(ibA_\pm) \, , \\
    \bm{\mathcal{M}}_2^\pm &= -\frac{i}{b A_\pm}
    \, \bm{Q}_\pm \,
    H_2^{(1)}(ib A_\pm) \, ,
\end{align}
\end{subequations}
where
\begin{equation}
    \bm{Z}_\pm = 
    \begin{pmatrix}
        ~1-A_\pm^2~ \\[3pt]
        -\mu A_\pm^2
    \end{pmatrix} , \quad
    \bm{Q}_\pm = 
    \begin{pmatrix}
        \mu A_\pm^2 \\[3pt]
        ~4 \left( 1-\xi^2 A_\pm^2 \right)~
    \end{pmatrix} .
\end{equation}
In addition, $\bm{\mathcal{C}}$ is a $2\times 3$ matrix with $\mathcal{C}_{23} = \rho/b^3$ and all other elements being zero.

\subsubsection{Solution outside the disk}

For $\rho \ge b$, representing the domain outside the disk, we obtain in complete analogy
\begin{equation}
    \bm{\mathcal{A}} = 
    \bm{\mathcal{N}}_+ \, H_1^{(1)}(i\rho A_+) - \bm{\mathcal{N}}_- \, H_1^{(1)}(i\rho A_-) \, ,
\end{equation}
where
\begin{equation}
     \bm{\mathcal{N}}_\pm =
     \frac{\pi}{8\varsigma^\frac{1}{2}}
     \left(
     \begin{matrix}
         ~\bm{\mathcal{N}}_1^\pm ~\large|~ \bm{\mathcal{N}}_2^\pm~
     \end{matrix}
     \right) ,
\end{equation}
with
\begin{subequations}
    \begin{align}
    \bm{\mathcal{N}}_1^\pm &= 
    \left(
       ~\bm{Z}_\pm ~ \large| ~ \bm{Q}_\pm~
    \right) \,
    I_1(b A_\pm) \, , \\
    \bm{\mathcal{N}}_2^\pm &= \frac{1}{b A_\pm}
    \, \bm{Q_\pm} \,  I_2(b A_\pm) \, .
\end{align}
\end{subequations}

It can be checked that for $\varsigma < 0$, 
$\bm{\mathcal{M}}_+$ and $\bm{\mathcal{M}}_-$ are negative complex conjugates and the same holds for $\bm{\mathcal{N}}_+$ and $\bm{\mathcal{N}}_-$.

\subsubsection{Solution for the force density}

To determine the three unknown force density coefficients, we require the boundary conditions to be satisfied at the surface of the rotating disk. This is accomplished by enforcing that $\bm{\mathcal{M}}_\pm \cdot \bm{\mathcal{X}} = \bm{0}$ and $\bm{\mathcal{C}} \cdot \bm{\mathcal{X}} = \left( 0, \rho/b \right)^\top$.
The latter equation implies that
\begin{equation}
    C = b^2 \, .
\end{equation}
We obtain
\begin{subequations}
    \begin{align}
    A &= \frac{\mu}{\varsigma^\frac{1}{2}}
    \left( \frac{2}{A_-^2}-\frac{2}{A_+^2} + \zeta_+ - \zeta_- \right), \\[3pt]
    B &= \frac{1}{\varsigma^\frac{1}{2}}
    \left( \frac{2-\zeta_+}{A_-^2}-\frac{2-\zeta_-}{A_+^2} + \zeta_+ - \zeta_- \right), 
\end{align}
\end{subequations}
where we have defined the abbreviation 
\begin{equation}
    \zeta_\pm = \frac{ib}{A_\pm} 
    \frac{H_0^{(1)} (ibA_\pm)}{H_1^{(1)}(ibA_\pm)} \, .
\end{equation}
Clearly, $A \propto \mu$ is the dominant term associated with odd viscosity, contributing to the force density acting on the perimeter of the disk along the radial direction.
It is worth noting that the coefficient $B$ also depends on $\mu$, albeit weakly, through $A_\pm$.

Finally, the radial and azimuthal velocities can be cast in a scaled form as
\begin{subequations} \label{eq:velocity_real_space_final}
    \begin{align}
   \frac{v_\rho}{a\Omega} &= \varphi_+ H_1^{(1)}(i\rho A_+) - \varphi_- H_1^{(1)}(i\rho A_-) \, , \label{eq:velocity_real_space_RADIAL} \\[5pt]
   \frac{v_\theta}{a\Omega} &= \frac{\rho}{b} + \psi_+ H_1^{(1)}(i\rho A_+) - \psi_- H_1^{(1)}(i\rho A_-) \, ,
\end{align}
\end{subequations}
where
\begin{equation}
    \left( \varphi_\pm, \psi_\pm \right)^\top
    = 
    \bm{\mathcal{N}}_\pm \cdot \bm{\mathcal{X}} \, .
\end{equation}
It can also be noticed that for $\varsigma < 0$, $\left( \varphi_+, \psi_+ \right)^\top$ and $\left( \varphi_-, \psi_- \right)^\top$ are negative complex conjugates. 

Since the expression for $K_n^q$, defined based on whether $\rho \leq b$ or $\rho \geq b$, is continuous at $\rho = b$, it can be readily verified at the disk surface that $\varphi_+ H_1^{(1)}(ib A_+) = \varphi_- H_1^{(1)}(ib A_-)$ and $\psi_+ H_1^{(1)}(ib A_+) = \psi_- H_1^{(1)}(ib A_-) $. Therefore, the boundary conditions are appropriately matched at the interface as required.

\begin{figure}
    \centering
    \includegraphics[width=0.9\linewidth]{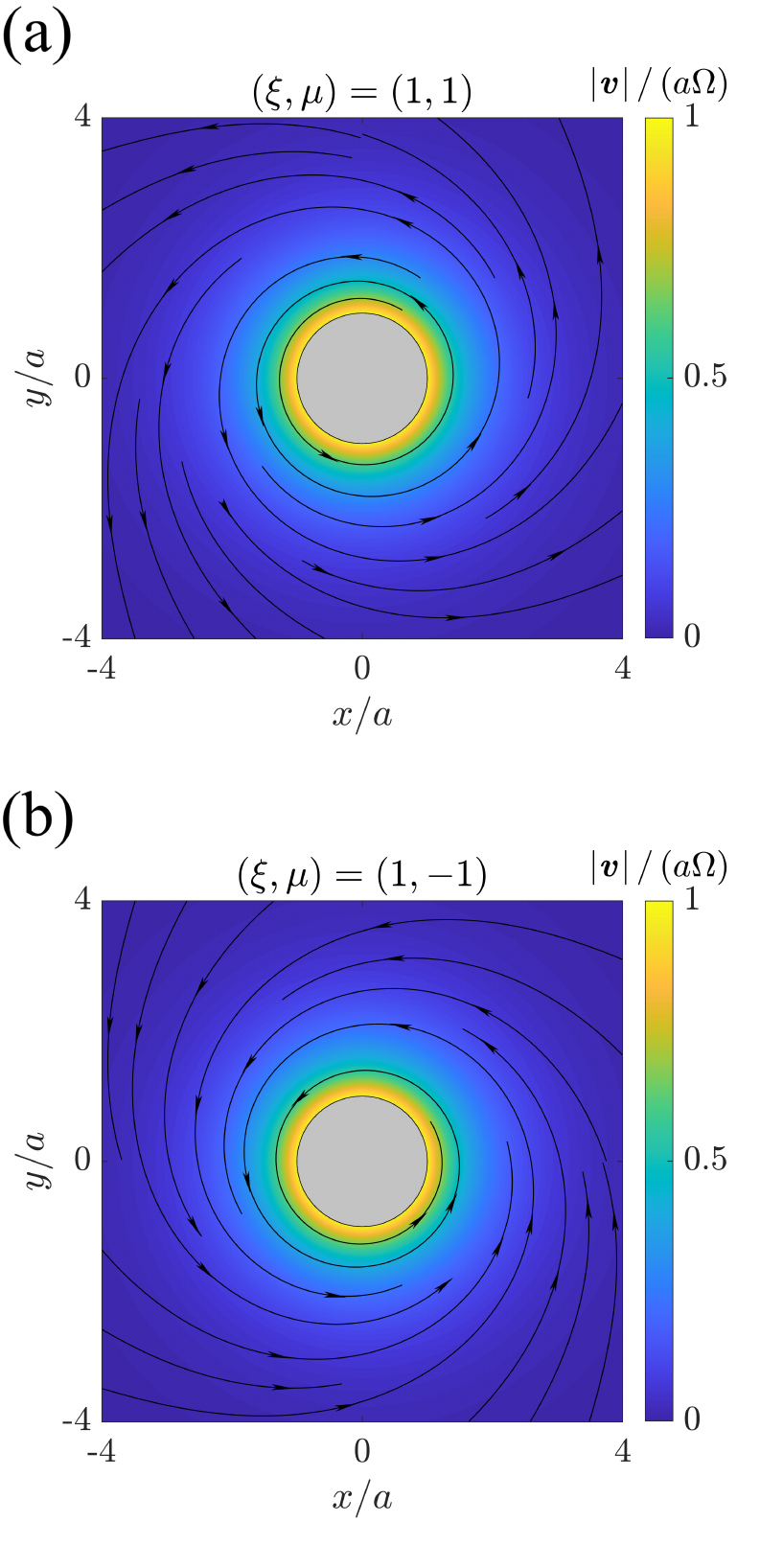} 
    \caption{(Color online) Streamlines and the magnitude of the scaled velocity field \( \bm{v} \) induced by the rotational motion of a disk rotating with steady angular velocity \( \Omega \), for (a) the odd-to-shear viscosity ratio \( \mu = \eta_\mathrm{O}/\eta_\mathrm{S} = 1 \) and (b) \( \mu = -1 \), while keeping \( \xi = 1 \) and \( b = 1 \). The disk is depicted in gray.}
    \label{fig:velocity_field}
\end{figure}

Figure~\ref{fig:velocity_field} shows exemplary directed streamlines and the flow field magnitude generated by a circular disk steadily spinning with angular velocity $\Omega$ in an odd fluid layer, with (a) $\mu = 1$ and (b) $\mu = -1$, while keeping $\xi = 1$ and $b = 1$. The flow is dominated by its azimuthal component, characterized by the swirling motion of fluid elements around the disk. The presence of odd viscosity introduces a radial component to the flow, directed outward or inward depending on the sign of the odd-to-shear viscosity ratio, $\mu$. A positive ratio results in a diverging vortical flow, where the fluid spirals outward, whereas a negative ratio leads to a converging vortical flow, with the fluid spiraling inward.

The emergence of normal components results from the non-reciprocal response induced by odd viscosity. In incompressible fluids, the swirling flow around rotating rigid objects is typically described by a classical rotlet singularity~\cite{happel2012low, chwang_wu_1974}, which implies that odd viscosity does not affect the flow field, as shown in both two-dimensional~\cite{avron1998, ganeshan2017} and three-dimensional~\cite{hosaka2024chirotactic} studies. 
This constraint imposed by the incompressible limit underscores the significance of the compressible regime, where odd viscosity plays a role in the flow around a rotating rigid object.

\subsection{Pressure field}

The solution for the pressure field can similarly be obtained via inverse Fourier transform of
\begin{equation}
    \widetilde{p} = \widetilde{\bm{\mathcal{P}}} \cdot \widetilde{\bm{f}} \, , 
\end{equation}
leading to another improper integral that can be evaluated exactly using the method of residues. However, a more straightforward approach is to solve Poisson's equation~\eqref{eq:compressibility} directly in real space, as we now have the solution for the velocity field. In polar coordinates, Eq.~\eqref{eq:compressibility} takes the form
\begin{equation}
    \frac{1}{\rho} \frac{\mathrm{d}}{\mathrm{d}\rho} \left( \rho v_\rho \right)
    = \frac{h^2}{6\eta_\mathrm{B}} \frac{1}{\rho}
    \frac{\mathrm{d}}{\mathrm{d}\rho} 
    \left( \rho \, \frac{\mathrm{d}p}{\mathrm{d}\rho} \right) . \label{eq:poisson}
\end{equation}

\begin{figure}
    \centering
    \includegraphics[width=0.9\linewidth]{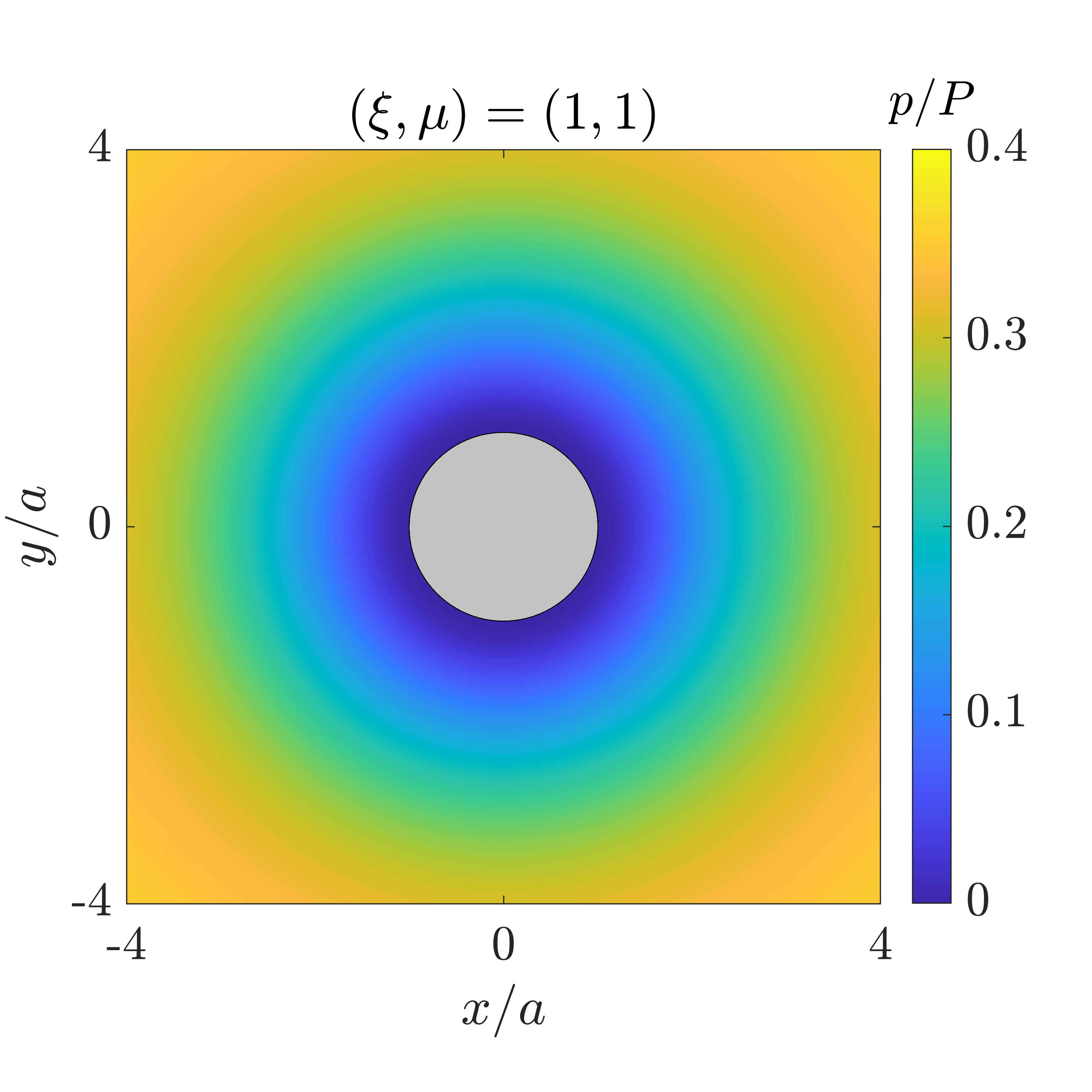}
    \caption{(Color online) Pressure field variations in the thin bulk fluid induced by the rotational motion of a disk spinning with a steady angular velocity \( \Omega \), for the odd-to-shear viscosity ratio \( \mu = \eta_\mathrm{O}/\eta_\mathrm{S} = 1 \), \( \xi = 1 \), and \( b = 1 \). The pressure is scaled by \( P = \Omega \, \eta_\mathrm{S} / h\).
    The disk is shown in gray.}
    \label{fig:pressure_field}
\end{figure}

By inserting the expression for the radial velocity from Eq.~\eqref{eq:velocity_real_space_RADIAL} into Eq.~\eqref{eq:poisson} and solving the resulting differential equation, the solution for the pressure field is obtained as
\begin{equation}
    \frac{p}{P} = 6ib \left( \frac{\varphi_+}{A_+} \, H_0^{(1)}(i\rho A_+)
    - \frac{\varphi_-}{A_-} \, H_0^{(1)}(i\rho A_-)
    \right) ,
\end{equation}
up to an additive constant, where $\Omega \, \eta_\mathrm{S} / h$ is used to rescale the pressure.
It follows that the pressure inside the disk is $p = p(\rho = b) = \text{constant}$. It can be verified that $\partial_\rho p = 0$ at the perimeter of the disk, where $\rho = b$.

Figure~\ref{fig:pressure_field} shows the variations in the pressure field around the spinning disk, with results presented for $\xi = \mu = b = 1$. In the absence of odd viscosity, the pressure is uniform across the fluid domain, as the flow is purely azimuthal. The presence of odd viscosity introduces a radial pressure distribution, which increases monotonically and eventually reaches a plateau at a distance far from the disk.
Such an odd viscosity-induced pressure due to a rotating object has been reported in incompressible fluids around a disk~\cite{avron1998} and a sphere~\cite{hosaka2024chirotactic}, and here we have extended it to compressible systems too.

\section{Rotational resistance coefficient}

Having derived exact analytical expressions for the hydrodynamic flow fields induced by a disk rotating steadily in a compressible fluid layer with odd viscosity on a rigid substrate, we now investigate the effect of the substrate-supported odd fluid layer on the rotational dynamics of a circular disk by calculating the rotational resistance coefficient. 

Using our framework, the resistance torque exerted on the disk by the surrounding fluid is related to the torque~$T$ averaged over the surface of the disk, $S$, via $\Lambda = -T/\Omega$.
The average torque acting on the disk along the~$z$-direction is expressed in terms of the azimuthal component of the force density as
\begin{equation}
    T = \frac{\Omega \, \eta_\mathrm{S}}{a} \int_S r f_\perp(r) \, \mathrm{d} S \, .
\end{equation}
Accordingly, the scaled resistance coefficient is obtained after integration as
\begin{equation}
    \frac{\Lambda}{2\pi a^2 \eta_\mathrm{S}} 
    = B + \frac{C}{4} \, .
\end{equation}
We further confirm that a rotating disk experiences no drag force, as required by the symmetry of the grand resistance tensor in chiral active fluids~\cite{khain2024trading}. In conjunction with the case of a translating disk examined earlier~\cite{daddi2025analytical}, we conclude that there is no translation-rotation coupling for the motion of a rigid disk in a fluid with odd viscosity.

In the limit $\mu \to 0$, we obtain
\begin{equation}
    \lim_{\mu \to 0} \, \frac{\Lambda}{2\pi a^2 \eta_\mathrm{S}} 
    = \frac{K_2(b)}{K_1(b)} \, b + \frac{b^2}{4}  \, ,
\end{equation}
with \( K_\nu \) denoting the modified Bessel function of the second kind of order \( \nu \)~\cite{abramowitz2000handbook}.
This results is in full agreement with the classical result by Evans and Sackmann~\cite{evans1988}, when the disk and substrate friction coefficients are set to be identical, i.e., $b_{\rm p} = b_{\rm s}$ in their notation.
We use $\Lambda_0$ to denote the resistance coefficient in this limit.

The resistance coefficient can be expressed as the sum of two contributions: \( \Lambda = \Lambda_\mathrm{S} + \Lambda_\mathrm{B} \), where \( \Lambda_\mathrm{S} \) is surface-related and \( \Lambda_\mathrm{B} \) is bulk-related.
The viscous torque exerted at perimeter is obtained as
\begin{equation}
    T_\mathrm{S} = -2\pi a^2 \sigma_{r\theta}(r=a) \, ,
\end{equation}
with the \( r\theta \)-component of the shear stress tensor in polar coordinates given by
\begin{equation}
    \sigma_{r\theta} = \eta_\mathrm{S}
    \left( \partial_r v_\theta - \frac{v_\theta}{r} - \mu \, \partial_r v_r \right) \, .
\end{equation}
This leads us to the surface-related contribution to the resistance coefficient, namely
\begin{equation}
    \frac{\Lambda_\mathrm{S}}{ 2\pi a^2 \eta_\mathrm{S}} = B \, .
\end{equation}

The viscous torque exerted on the bottom of the disk is the torque exerted by the bulk force \( \bm{f}_\mathrm{B} = f_\mathrm{B} \, \hat{\bm{e}}_\theta \), where \( f_\mathrm{B} = -\kappa^2 \eta_\mathrm{S} \Omega \, r \). Accordingly, the bulk-related contribution to the resistance coefficient is obtained as
\begin{equation}
    \frac{\Lambda_\mathrm{B}}{2\pi a^2 \eta_\mathrm{S}} = \frac{C}{4}  \, .
\end{equation}
We note that $\eta_\mathrm{B}=\kappa^2 h  \eta_\mathrm{S}$.

\begin{figure}
    \centering
    \includegraphics[width=0.9\linewidth]{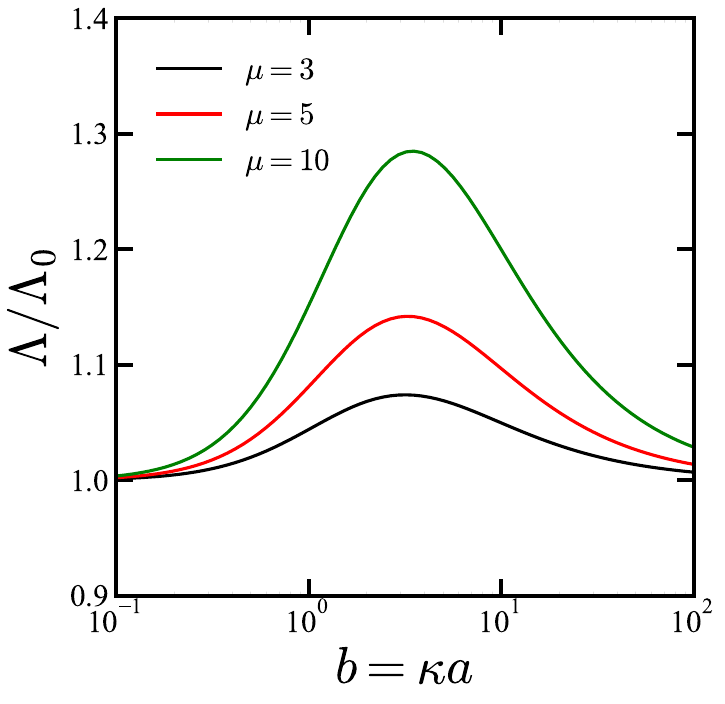} 
    \caption{ (Color online) The variation of the relative rotational resistance coefficient \( \Lambda/\Lambda_0 \) as a function of the rescaled disk radius \( b = \kappa a \) for different values of the odd-to-even viscosity ratio \( \mu = \eta_{\rm O}/\eta_{\rm S} \), with \( \xi = 1 \) held constant. In the plot, \( \mu = 3 \), 5, and 10 are represented by the solid black, red, and green lines, respectively.
    }
    \label{fig:resistance}
\end{figure}

In either approach, we obtain identical expressions for the rotational resistance coefficient, demonstrating the correctness and validity of our method.
Figure~\ref{fig:resistance} shows the relative resistance coefficient \( \Lambda/\Lambda_0 \) as a function of the rescaled radius \( b = \kappa a \) for various values of the odd-to-even viscosity ratio \( \mu = \eta_{\rm O}/\eta_{\rm S} \). 
The curves exhibit a bell-shaped behavior, peaking around \( b \sim 5 \), with \( \Lambda \sim \Lambda_0 \) both for \( b \ll 1 \) and \( b \gg 1 \). Increasing~\( \mu \) results in a higher~\( \Lambda \), which corresponds to an increase in the rotational friction of the disk.

In the case, when \( b \ll 1 \), we obtain
\begin{equation}
    \frac{\Lambda}{2\pi a^2 \eta_\mathrm{S}}  = 
    2 + \left( \frac{1}{4}+\ln 2-\gamma + \frac{H}{\varsigma^\frac{1}{2}} \right)b^2 +
    \mathcal{O} \left( b^4 \right) , 
\end{equation}
with \( \gamma \approx 0.5772 \ldots \) denoting the Euler-Mascheroni constant and 
\begin{equation}
    H = \ln \left( \frac{A_+}{A_-} \right) + \frac{\ln \left( b A_-\right)}{A_+^2} - \frac{\ln \left( b A_+\right)}{A_-^2} \, .
\end{equation}
In the opposite limit, where \( b \gg 1 \), we obtain
\begin{equation}
    \frac{\Lambda}{2\pi a^2 \eta_\mathrm{S}} = 
    \frac{b^2}{4} + \mathcal{O} \left( b\right) .
\end{equation}
Accordingly, in the regime of large values of~\( b \), the resistance coefficient is primarily determined by the bulk contribution, which plays the dominant role.

In both cases, the effect of odd viscosity on the drag coefficient is of quadratic order. 
The dependence reflects the fact that 2D odd viscosity on its own is dissipationless and only contributes to drag via its effect on the flow field.
A dependence of even order can also be expected on grounds of symmetry, with dissipation being a scalar and odd viscosity a pseudoscalar. Another way of reasoning is linked to the Helmholtz minimum dissipation theorem \cite{guazzelli2011physical}, which states that any perturbation to the pure Stokes flow, while keeping the velocity boundary conditions fixed, increases the viscous dissipation. 

Note that these results are specific to compressible fluids. In a 2D incompressible fluid, odd viscosity has no effect on rotational drag of a rigid disk~\cite{avron1998, lapa2014, ganeshan2017}. In 3D, it affects the drag, but the contribution of odd viscosity is perpendicular to angular velocity and therefore does not contribute to dissipation \cite{everts2024dissipative, hosaka2024chirotactic}.

\section{Conclusion}
In conclusion, we have analyzed the hydrodynamics of a rotating disk in a 2D compressible fluid layer with odd viscosity. We have demonstrated that the emergence of a diverging spiral flow around a rotating disk is a distinctive feature of odd viscosity. 
The radial flow component associated with this behavior can be directed inward or outward, depending on the sign of the odd viscosity. The interplay between odd viscosity and rotation can in principle be expected on the basis of fundamental symmetries. 
In a 2D system, angular velocity and odd viscosity are both pseudoscalars (pseudovectors in 3D). 
Their product, a scalar, then decides between repulsion if $\Omega$ and $\eta_{\rm O}$ have the same sign or attraction if they have opposite signs.
This effect has significant implications for the hydrodynamic interactions between multiple disks. Specifically, the sign of the odd viscosity determines whether two rotating disks experience effective attraction or repulsion, along with the generation of pairwise torques. 
Finally, using analytic solutions for the flow and pressure fields, we compute the rotational drag of the rotating disk. We show that odd viscosity increases rotational drag, providing a viable experimental method for detecting odd viscosity.

\begin{acknowledgments}
A.V. acknowledges support from the Slovenian Research and Innovation Agency (Grants No.\ P1-0099 and J1-60009).
E.T. acknowledges funding from EPSRC (Grant No.\ EP/W027194/1).
\end{acknowledgments}

\section*{Data Availability Statement}
The data that support the findings of this study are available from the corresponding author upon reasonable request. 

\appendix*

\section{Analytical evaluation of $Z_n^q$}

In this Appendix, we provide detailed derivations related to the evaluation of the series improper integral~$Z_n^q$ defined by Eq.~\eqref{eq:Znq}.

We define the following function in the complex plane
\begin{equation}
    f(z) = 
    \begin{cases}
        \cfrac{z^{q} H_{n}^{(1)}(bz) J_{1}({\rho} z)}{4 \left( z^2+1 \right) 
    \left( \xi^2 z^2+1\right) + \mu^2 z^4} & \text{if } {\rho} \le b \, , \\[5pt]
        \cfrac{z^{q} J_{n}(bz) H_{1}^{(1)}({\rho} z)}{4 \left( z^2+1 \right) 
        \left( \xi^2 z^2+1\right) + \mu^2 z^4}  & \text{if } {\rho} \ge b \, , 
    \end{cases}
    \label{eq:f_von_z}
\end{equation}
with $H_n^{(1)}$ standing for the $n$th-order Hankel function of the first kind~\cite{abramowitz2000handbook}.
For the relevant values of the pair $(n, q)$, the integral is convergent. Since $n + q$ is even, the real part of $f(z)$ is even, and the imaginary part is odd. Therefore, we have
\begin{equation}
    Z_n^q = \frac{1}{2} \dashint_{-\infty}^\infty f(x) \, \mathrm{d} x \, ,
\end{equation}
where the integral is understood in the sense of the Cauchy principal value, denoted by a dash.

The function $f(z)$ is meromorphic because it is analytic everywhere except at a finite number of poles.
We integrate the function \( f(z) \) in the complex plane along a closed contour~$\mathcal{C}$ consisting of the following segments: the upper part of the branch cut from \( -R \) to \( -\epsilon \), with~$R$ and~$\epsilon$ being positive numbers, where $\epsilon \ll R$, a small clockwise semicircle of radius \( \epsilon \) centered at the origin of coordinates denoted by $\mathcal{C}_\epsilon$, the positive real axis from \( \epsilon \) to~\( R \), and the counterclockwise contour along the upper half of the large semicircle \( |z| = R \) denoted by $\mathcal{C}_R$. The choice of \( f(z) \) depends on whether \( \rho \leq b \) or \( \rho \geq b \), ensuring that the integral over the large semicircle vanishes for an infinitely large \( R \), as detailed below. By the residue theorem, the integral along the closed contour is expressed as
\begin{equation}
    \int_\mathcal{C} f(z) \, \mathrm{d}z = 2i\pi \sum_j \text{Res} \{ f(z), z=z_j\} \, , 
\end{equation}
where $z_j$ denotes the poles of $f(z)$, located at \( z = iA_\pm \), where
\begin{equation}
    A_\pm = \left( \frac{2 \left( 1+\xi^2 \pm \varsigma^{\frac{1}{2}} \right)}{\mu^2 + 4 \xi^2} \right)^\frac{1}{2} ,
\end{equation}
where $\varsigma = \left( 1-\xi^2\right)^2-\mu^2$. The imaginary part of \( z = iA_\pm \) is always positive, positioning the poles in the upper half of the complex plane. When \( \varsigma \geq 0 \), \( z = iA_\pm \) are purely imaginary, with $A_+ \ge A_-$, whereas for \( \varsigma < 0 \), \( A_+ \) and~\( A_- \) are complex conjugates.

For vanishing odd viscosity, where $\mu = 0$, it follows that $(A_+, A_- ) = (1/\xi, 1)$ for $\xi \in [1/2,1]$, and $(A_+, A_- ) = (1, 1/\xi)$ for $\xi \ge 1$.

Thus, the contour integral over the three segments leads us to
\begin{align}
    \dashint_{-\infty}^\infty f(x) \, \mathrm{d}x
    + \lim_{\epsilon\to 0} \int_{\mathcal{C}_\epsilon} f(z) \, \mathrm{d}z
    + \lim_{R\to\infty} \int_{\mathcal{C}_R} f(z) \, \mathrm{d}z = \notag \\[3pt]
    2i\pi \big( \operatorname{Res} \{f(z),z=iA_+\} + \operatorname{Res} \{f(z),z=iA_-\} \big) \, .
\end{align}

The residues at $z = iA_\pm$ are obtained as
\begin{align}
    \operatorname{Res}  \{ f(z), z &= iA_\pm\} = 
    \pm\frac{\pi}{8 \varsigma^\frac{1}{2}} \, (iA_\pm)^{q-1} \, \times\notag \\
    & \begin{cases}
         H_n(ibA_\pm) \, I_1(\rho A_\pm) \qquad \,\,\text{if~} \rho \le b \, , \\[3pt]
        i^{n-1} I_n(bA_\pm) \,H_1(i\rho A_\pm ) \,\,\, \text{if~} \rho \ge b \, .
    \end{cases}
\end{align}

To evaluate the contour integral over the large semicircle defined by $\mathcal{C}_R = \{ R\,e^{i\theta} \mid \theta \in [0,\pi] \}$, we use Jordan's lemma~\cite{brown2009complex}. 
The lemma states that if a complex function is of the form
\begin{equation}
f(z) = e^{i\tau z} g(z) \, ,
\end{equation}
with $\tau>0$, then the following upper bound for the contour integral holds:
\begin{equation}
\left| \int_{\mathcal{C}_R} f(z) \, \mathrm{d}z \right| \le \frac{\pi}{\tau} \, M \, , 
\label{eq:jordan}
\end{equation}
where
\begin{equation}
M = \max_{\theta \in [0,\pi]} \left| g \left( R e^{i\theta} \right) \right|.
\end{equation}
Equality in Eq.~\eqref{eq:jordan} holds when $g$ vanishes everywhere, in which case both sides are identically zero. 

When $\tau = 0$, we use the estimation lemma, also known as the ML inequality, to show that the integral vanishes.
The estimation lemma provides an upper bound for a contour integral, stating that if $|f(z)|$ is bounded by a constant~$M$ for all~$z$ on a contour $\mathcal{C}$, then
\begin{equation}
    \left| \int_{\mathcal{C}} f(z) \, \mathrm{d}z \right| \le M \ell(\mathcal{C}) \, ,
\end{equation}
where $\ell(\mathcal{C})$ represents the arc length of $\mathcal{C}$. In particular, we may take the maximum
\begin{equation}
    M = \sup_{z \in \mathcal{C}} |f(z)| \, .
\end{equation}

Thus, if $f$ is continuous on the semicircular contour~$\mathcal{C}_R$ for all large~$R$ and $\lim_{R\to \infty} M = 0$, it follows from these lemmas that the integral over the semicircle vanishes.

To apply these lemmas, we examine the behavior of~$f(z)$ for an infinitely large~$z$.
As $z \to \infty$, the asymptotic behavior is given by
\begin{equation}
    f(z) \sim \frac{2N}{z^{5-q}}
    \begin{cases}
        \cos \left( \rho z+\frac\pi4 \right) e^{i \left( bz - (2n+1) \frac\pi4 \right)} & \text{if } \rho \le b \\[3pt]
        \sin \left( bz - (2n-1) \frac{\pi}{4} \right) e^{i \left( \rho z + \frac\pi4\right)} & \text{if } \rho \ge b 
    \end{cases}
\end{equation}
with 
\begin{equation}
    N^{-1} = -\pi \left( \rho b\right)^\frac{1}{2} \left( \mu^2+4\xi^2\right) \, .
\end{equation}

By expressing cosine and sine in their exponential forms using Euler's relations, we obtain 
\begin{equation}
	\hspace{-0.25cm}
    f(z) \sim \frac{N}{z^{5-q}}
    \left( e^{i \left( (\rho+b)z-n \, \frac{\pi}{2} \right)} 
    + e^{i \left( |\rho-b|z+\varepsilon (n+1) \frac{\pi}{2} \right)}
    \right)
\end{equation}
where $\varepsilon = \operatorname{sgn}(\rho - b)$, with sgn representing the sign function.

For the first integral, we apply Jordan's lemma by setting $\tau = \rho + b > 0$ to demonstrate that it vanishes over the large semicircle for all strictly positive values of $\rho$ and $b$. For the second integral, we set $\tau = |\rho - b| \ge 0$ and use Jordan's lemma to show that it vanishes when $\rho \neq b$. When $\rho = b$, the second integral vanishes by the estimation lemma.
Thus, the integral over $\mathcal{C}_R$ vanishes as $R \to \infty$.

Now, let us concentrate on the contour integral along the small, clockwise-oriented semicircle of radius~$\epsilon$ centered at the origin, denoted by $\mathcal{C}_\epsilon = \{ \epsilon e^{i\theta} \mid \theta \in [\pi, 0] \}$.
We can apply the residue theorem by considering in the lower plane a complementary semicircle of radius $\epsilon$.
This method leads to
\begin{equation}
    \int_{\mathcal{C}_\epsilon} f(z) \, \mathrm{d} z
    = -i\pi \operatorname{Res} \{f(z),z=0\} \, , 
\end{equation}
where the negative sign results from the clockwise orientation of the contour integration.
The residue at $z = 0$ is equal to the coefficient of the $z^{-1}$ term in the principal part of the Laurent series expansion of $f(z)$ around $z = 0$. We find that this term vanishes except for $(n, q) = (2, 0)$ when $\rho \leq b$, as given by $K_n^q$ defined in Eq.~\eqref{eq:Knq}.

\section*{References}


\begin{thebibliography}{62}%
\makeatletter
\providecommand \@ifxundefined [1]{%
 \@ifx{#1\undefined}
}%
\providecommand \@ifnum [1]{%
 \ifnum #1\expandafter \@firstoftwo
 \else \expandafter \@secondoftwo
 \fi
}%
\providecommand \@ifx [1]{%
 \ifx #1\expandafter \@firstoftwo
 \else \expandafter \@secondoftwo
 \fi
}%
\providecommand \natexlab [1]{#1}%
\providecommand \enquote  [1]{``#1''}%
\providecommand \bibnamefont  [1]{#1}%
\providecommand \bibfnamefont [1]{#1}%
\providecommand \citenamefont [1]{#1}%
\providecommand \href@noop [0]{\@secondoftwo}%
\providecommand \href [0]{\begingroup \@sanitize@url \@href}%
\providecommand \@href[1]{\@@startlink{#1}\@@href}%
\providecommand \@@href[1]{\endgroup#1\@@endlink}%
\providecommand \@sanitize@url [0]{\catcode `\\12\catcode `\$12\catcode
  `\&12\catcode `\#12\catcode `\^12\catcode `\_12\catcode `\%12\relax}%
\providecommand \@@startlink[1]{}%
\providecommand \@@endlink[0]{}%
\providecommand \url  [0]{\begingroup\@sanitize@url \@url }%
\providecommand \@url [1]{\endgroup\@href {#1}{\urlprefix }}%
\providecommand \urlprefix  [0]{URL }%
\providecommand \Eprint [0]{\href }%
\providecommand \doibase [0]{https://doi.org/}%
\providecommand \selectlanguage [0]{\@gobble}%
\providecommand \bibinfo  [0]{\@secondoftwo}%
\providecommand \bibfield  [0]{\@secondoftwo}%
\providecommand \translation [1]{[#1]}%
\providecommand \BibitemOpen [0]{}%
\providecommand \bibitemStop [0]{}%
\providecommand \bibitemNoStop [0]{.\EOS\space}%
\providecommand \EOS [0]{\spacefactor3000\relax}%
\providecommand \BibitemShut  [1]{\csname bibitem#1\endcsname}%
\let\auto@bib@innerbib\@empty
\bibitem [{\citenamefont {Shankar}\ \emph {et~al.}(2022)\citenamefont
  {Shankar}, \citenamefont {Souslov}, \citenamefont {Bowick}, \citenamefont
  {Marchetti},\ and\ \citenamefont {Vitelli}}]{shankar2022topological}%
  \BibitemOpen
  \bibfield  {author} {\bibinfo {author} {\bibfnamefont {S.}~\bibnamefont
  {Shankar}}, \bibinfo {author} {\bibfnamefont {A.}~\bibnamefont {Souslov}},
  \bibinfo {author} {\bibfnamefont {M.~J.}\ \bibnamefont {Bowick}}, \bibinfo
  {author} {\bibfnamefont {M.~C.}\ \bibnamefont {Marchetti}},\ and\ \bibinfo
  {author} {\bibfnamefont {V.}~\bibnamefont {Vitelli}},\ }\bibfield  {title}
  {\bibinfo {title} {Topological active matter},\ }\href
  {https://doi.org/10.1038/s42254-022-00445-3} {\bibfield  {journal} {\bibinfo
  {journal} {Nat. Rev. Phys.}\ }\textbf {\bibinfo {volume} {4}},\ \bibinfo
  {pages} {380} (\bibinfo {year} {2022})}\BibitemShut {NoStop}%
\bibitem [{\citenamefont {Oppenheimer}\ \emph {et~al.}(2022)\citenamefont
  {Oppenheimer}, \citenamefont {Stein}, \citenamefont {Zion},\ and\
  \citenamefont {Shelley}}]{oppenheimer2022hyperuniformity}%
  \BibitemOpen
  \bibfield  {author} {\bibinfo {author} {\bibfnamefont {N.}~\bibnamefont
  {Oppenheimer}}, \bibinfo {author} {\bibfnamefont {D.~B.}\ \bibnamefont
  {Stein}}, \bibinfo {author} {\bibfnamefont {M.~Y.~B.}\ \bibnamefont {Zion}},\
  and\ \bibinfo {author} {\bibfnamefont {M.~J.}\ \bibnamefont {Shelley}},\
  }\bibfield  {title} {\bibinfo {title} {Hyperuniformity and phase enrichment
  in vortex and rotor assemblies},\ }\href
  {https://doi.org/https://doi.org/10.1038/s41467-022-28375-9} {\bibfield
  {journal} {\bibinfo  {journal} {Nat. Commun.}\ }\textbf {\bibinfo {volume}
  {13}},\ \bibinfo {pages} {804} (\bibinfo {year} {2022})}\BibitemShut
  {NoStop}%
\bibitem [{\citenamefont {Drescher}\ \emph {et~al.}(2009)\citenamefont
  {Drescher}, \citenamefont {Leptos}, \citenamefont {Tuval}, \citenamefont
  {Ishikawa}, \citenamefont {Pedley},\ and\ \citenamefont
  {Goldstein}}]{drescher2009dancing}%
  \BibitemOpen
  \bibfield  {author} {\bibinfo {author} {\bibfnamefont {K.}~\bibnamefont
  {Drescher}}, \bibinfo {author} {\bibfnamefont {K.~C.}\ \bibnamefont
  {Leptos}}, \bibinfo {author} {\bibfnamefont {I.}~\bibnamefont {Tuval}},
  \bibinfo {author} {\bibfnamefont {T.}~\bibnamefont {Ishikawa}}, \bibinfo
  {author} {\bibfnamefont {T.~J.}\ \bibnamefont {Pedley}},\ and\ \bibinfo
  {author} {\bibfnamefont {R.~E.}\ \bibnamefont {Goldstein}},\ }\bibfield
  {title} {\bibinfo {title} {Dancing {V}olvox: Hydrodynamic bound states of
  swimming algae},\ }\href {https://doi.org/10.1103/PhysRevLett.102.168101}
  {\bibfield  {journal} {\bibinfo  {journal} {Phys. Rev. Lett.}\ }\textbf
  {\bibinfo {volume} {102}},\ \bibinfo {pages} {168101} (\bibinfo {year}
  {2009})}\BibitemShut {NoStop}%
\bibitem [{\citenamefont {Tan}\ \emph {et~al.}(2022)\citenamefont {Tan},
  \citenamefont {Mietke}, \citenamefont {Li}, \citenamefont {Chen},
  \citenamefont {Higinbotham}, \citenamefont {Foster}, \citenamefont {Gokhale},
  \citenamefont {Dunkel},\ and\ \citenamefont {Fakhri}}]{tan2022odd}%
  \BibitemOpen
  \bibfield  {author} {\bibinfo {author} {\bibfnamefont {T.~H.}\ \bibnamefont
  {Tan}}, \bibinfo {author} {\bibfnamefont {A.}~\bibnamefont {Mietke}},
  \bibinfo {author} {\bibfnamefont {J.}~\bibnamefont {Li}}, \bibinfo {author}
  {\bibfnamefont {Y.}~\bibnamefont {Chen}}, \bibinfo {author} {\bibfnamefont
  {H.}~\bibnamefont {Higinbotham}}, \bibinfo {author} {\bibfnamefont {P.~J.}\
  \bibnamefont {Foster}}, \bibinfo {author} {\bibfnamefont {S.}~\bibnamefont
  {Gokhale}}, \bibinfo {author} {\bibfnamefont {J.}~\bibnamefont {Dunkel}},\
  and\ \bibinfo {author} {\bibfnamefont {N.}~\bibnamefont {Fakhri}},\
  }\bibfield  {title} {\bibinfo {title} {Odd dynamics of living chiral
  crystals},\ }\href {https://doi.org/10.1038/s41586-022-04889-6} {\bibfield
  {journal} {\bibinfo  {journal} {Nature}\ }\textbf {\bibinfo {volume} {607}},\
  \bibinfo {pages} {287} (\bibinfo {year} {2022})}\BibitemShut {NoStop}%
\bibitem [{\citenamefont {Lauga}\ \emph {et~al.}(2006)\citenamefont {Lauga},
  \citenamefont {DiLuzio}, \citenamefont {Whitesides},\ and\ \citenamefont
  {Stone}}]{lauga2006swimming}%
  \BibitemOpen
  \bibfield  {author} {\bibinfo {author} {\bibfnamefont {E.}~\bibnamefont
  {Lauga}}, \bibinfo {author} {\bibfnamefont {W.~R.}\ \bibnamefont {DiLuzio}},
  \bibinfo {author} {\bibfnamefont {G.~M.}\ \bibnamefont {Whitesides}},\ and\
  \bibinfo {author} {\bibfnamefont {H.~A.}\ \bibnamefont {Stone}},\ }\bibfield
  {title} {\bibinfo {title} {Swimming in circles: motion of bacteria near solid
  boundaries},\ }\href {https://doi.org/10.1529/biophysj.105.069401} {\bibfield
   {journal} {\bibinfo  {journal} {Biophys. J.}\ }\textbf {\bibinfo {volume}
  {90}},\ \bibinfo {pages} {400} (\bibinfo {year} {2006})}\BibitemShut
  {NoStop}%
\bibitem [{\citenamefont {Yang}\ \emph {et~al.}(2020)\citenamefont {Yang},
  \citenamefont {Ren}, \citenamefont {Cheng},\ and\ \citenamefont
  {Zhang}}]{yang2020}%
  \BibitemOpen
  \bibfield  {author} {\bibinfo {author} {\bibfnamefont {X.}~\bibnamefont
  {Yang}}, \bibinfo {author} {\bibfnamefont {C.}~\bibnamefont {Ren}}, \bibinfo
  {author} {\bibfnamefont {K.}~\bibnamefont {Cheng}},\ and\ \bibinfo {author}
  {\bibfnamefont {H.}~\bibnamefont {Zhang}},\ }\bibfield  {title} {\bibinfo
  {title} {Robust boundary flow in chiral active fluid},\ }\href
  {https://doi.org/https://doi.org/10.1103/PhysRevE.101.022603} {\bibfield
  {journal} {\bibinfo  {journal} {Phys. Rev. E}\ }\textbf {\bibinfo {volume}
  {101}},\ \bibinfo {pages} {022603} (\bibinfo {year} {2020})}\BibitemShut
  {NoStop}%
\bibitem [{\citenamefont {Soni}\ \emph {et~al.}(2019)\citenamefont {Soni},
  \citenamefont {Bililign}, \citenamefont {Magkiriadou}, \citenamefont
  {Sacanna}, \citenamefont {Bartolo}, \citenamefont {Shelley},\ and\
  \citenamefont {Irvine}}]{soni2019odd}%
  \BibitemOpen
  \bibfield  {author} {\bibinfo {author} {\bibfnamefont {V.}~\bibnamefont
  {Soni}}, \bibinfo {author} {\bibfnamefont {E.~S.}\ \bibnamefont {Bililign}},
  \bibinfo {author} {\bibfnamefont {S.}~\bibnamefont {Magkiriadou}}, \bibinfo
  {author} {\bibfnamefont {S.}~\bibnamefont {Sacanna}}, \bibinfo {author}
  {\bibfnamefont {D.}~\bibnamefont {Bartolo}}, \bibinfo {author} {\bibfnamefont
  {M.~J.}\ \bibnamefont {Shelley}},\ and\ \bibinfo {author} {\bibfnamefont
  {W.~T.~M.}\ \bibnamefont {Irvine}},\ }\bibfield  {title} {\bibinfo {title}
  {The odd free surface flows of a colloidal chiral fluid},\ }\href
  {https://doi.org/10.1038/s41567-019-0603-8} {\bibfield  {journal} {\bibinfo
  {journal} {Nat. Phys.}\ }\textbf {\bibinfo {volume} {15}},\ \bibinfo {pages}
  {1188} (\bibinfo {year} {2019})}\BibitemShut {NoStop}%
\bibitem [{\citenamefont {Mecke}\ \emph
  {et~al.}(2024{\natexlab{a}})\citenamefont {Mecke}, \citenamefont {Gao},
  \citenamefont {Gompper},\ and\ \citenamefont {Ripoll}}]{mecke2024chiral}%
  \BibitemOpen
  \bibfield  {author} {\bibinfo {author} {\bibfnamefont {J.}~\bibnamefont
  {Mecke}}, \bibinfo {author} {\bibfnamefont {Y.}~\bibnamefont {Gao}}, \bibinfo
  {author} {\bibfnamefont {G.}~\bibnamefont {Gompper}},\ and\ \bibinfo {author}
  {\bibfnamefont {M.}~\bibnamefont {Ripoll}},\ }\bibfield  {title} {\bibinfo
  {title} {Chiral active systems near a substrate: Emergent damping length
  controlled by fluid friction},\ }\href
  {https://doi.org/https://doi.org/10.1038/s42005-024-01817-0} {\bibfield
  {journal} {\bibinfo  {journal} {Commun. Phys.}\ }\textbf {\bibinfo {volume}
  {7}},\ \bibinfo {pages} {332} (\bibinfo {year}
  {2024}{\natexlab{a}})}\BibitemShut {NoStop}%
\bibitem [{\citenamefont {Onsager}(1931)}]{onsager1931reciprocal}%
  \BibitemOpen
  \bibfield  {author} {\bibinfo {author} {\bibfnamefont {L.}~\bibnamefont
  {Onsager}},\ }\bibfield  {title} {\bibinfo {title} {Reciprocal relations in
  irreversible processes. {II.}},\ }\href
  {https://doi.org/https://doi.org/10.1103/PhysRev.38.2265} {\bibfield
  {journal} {\bibinfo  {journal} {Phys. Rev.}\ }\textbf {\bibinfo {volume}
  {38}},\ \bibinfo {pages} {2265} (\bibinfo {year} {1931})}\BibitemShut
  {NoStop}%
\bibitem [{\citenamefont {Masoud}\ and\ \citenamefont
  {Stone}(2019)}]{masoud2019}%
  \BibitemOpen
  \bibfield  {author} {\bibinfo {author} {\bibfnamefont {H.}~\bibnamefont
  {Masoud}}\ and\ \bibinfo {author} {\bibfnamefont {H.~A.}\ \bibnamefont
  {Stone}},\ }\bibfield  {title} {\bibinfo {title} {The reciprocal theorem in
  fluid dynamics and transport phenomena},\ }\href
  {https://doi.org/10.1017/jfm.2019.553} {\bibfield  {journal} {\bibinfo
  {journal} {J. Fluid Mech.}\ }\textbf {\bibinfo {volume} {879}},\ \bibinfo
  {pages} {P1} (\bibinfo {year} {2019})}\BibitemShut {NoStop}%
\bibitem [{\citenamefont {Doi}(2013)}]{doi2013soft}%
  \BibitemOpen
  \bibfield  {author} {\bibinfo {author} {\bibfnamefont {M.}~\bibnamefont
  {Doi}},\ }\href
  {https://doi.org/https://doi.org/10.1093/acprof:oso/9780199652952.001.0001}
  {\emph {\bibinfo {title} {Soft Matter Physics}}}\ (\bibinfo  {publisher}
  {Oxford University Press},\ \bibinfo {address} {New York},\ \bibinfo {year}
  {2013})\BibitemShut {NoStop}%
\bibitem [{\citenamefont {Avron}(1998)}]{avron1998}%
  \BibitemOpen
  \bibfield  {author} {\bibinfo {author} {\bibfnamefont {J.~E.}\ \bibnamefont
  {Avron}},\ }\bibfield  {title} {\bibinfo {title} {Odd viscosity},\ }\href
  {https://doi.org/10.1023/A:1023084404080} {\bibfield  {journal} {\bibinfo
  {journal} {J. Stat. Phys.}\ }\textbf {\bibinfo {volume} {92}},\ \bibinfo
  {pages} {543} (\bibinfo {year} {1998})}\BibitemShut {NoStop}%
\bibitem [{\citenamefont {Banerjee}\ \emph {et~al.}(2017)\citenamefont
  {Banerjee}, \citenamefont {Souslov}, \citenamefont {Abanov},\ and\
  \citenamefont {Vitelli}}]{banerjee2017}%
  \BibitemOpen
  \bibfield  {author} {\bibinfo {author} {\bibfnamefont {D.}~\bibnamefont
  {Banerjee}}, \bibinfo {author} {\bibfnamefont {A.}~\bibnamefont {Souslov}},
  \bibinfo {author} {\bibfnamefont {A.~G.}\ \bibnamefont {Abanov}},\ and\
  \bibinfo {author} {\bibfnamefont {V.}~\bibnamefont {Vitelli}},\ }\bibfield
  {title} {\bibinfo {title} {Odd viscosity in chiral active fluids},\ }\href
  {https://doi.org/10.1038/s41467-017-01378-7} {\bibfield  {journal} {\bibinfo
  {journal} {Nat. Commun.}\ }\textbf {\bibinfo {volume} {8}},\ \bibinfo {pages}
  {1573} (\bibinfo {year} {2017})}\BibitemShut {NoStop}%
\bibitem [{\citenamefont {Hosaka}\ and\ \citenamefont
  {Komura}(2022)}]{hosaka2022nonequilibrium}%
  \BibitemOpen
  \bibfield  {author} {\bibinfo {author} {\bibfnamefont {Y.}~\bibnamefont
  {Hosaka}}\ and\ \bibinfo {author} {\bibfnamefont {S.}~\bibnamefont
  {Komura}},\ }\bibfield  {title} {\bibinfo {title} {Nonequilibrium transport
  induced by biological nanomachines},\ }\href
  {https://doi.org/https://doi.org/10.1142/S1793048022310026} {\bibfield
  {journal} {\bibinfo  {journal} {Biophys. Rev. Lett.}\ }\textbf {\bibinfo
  {volume} {17}},\ \bibinfo {pages} {51} (\bibinfo {year} {2022})}\BibitemShut
  {NoStop}%
\bibitem [{\citenamefont {Fruchart}\ \emph {et~al.}(2023)\citenamefont
  {Fruchart}, \citenamefont {Scheibner},\ and\ \citenamefont
  {Vitelli}}]{fruchart2023odd}%
  \BibitemOpen
  \bibfield  {author} {\bibinfo {author} {\bibfnamefont {M.}~\bibnamefont
  {Fruchart}}, \bibinfo {author} {\bibfnamefont {C.}~\bibnamefont
  {Scheibner}},\ and\ \bibinfo {author} {\bibfnamefont {V.}~\bibnamefont
  {Vitelli}},\ }\bibfield  {title} {\bibinfo {title} {Odd viscosity and odd
  elasticity},\ }\href
  {https://doi.org/10.1146/annurev-conmatphys-040821-125506} {\bibfield
  {journal} {\bibinfo  {journal} {Annu. Rev. Condens. Matter Phys.}\ }\textbf
  {\bibinfo {volume} {14}},\ \bibinfo {pages} {471} (\bibinfo {year}
  {2023})}\BibitemShut {NoStop}%
\bibitem [{\citenamefont {Mecke}\ \emph
  {et~al.}(2024{\natexlab{b}})\citenamefont {Mecke}, \citenamefont {Nketsiah},
  \citenamefont {Li},\ and\ \citenamefont {Gao}}]{mecke2024emergent}%
  \BibitemOpen
  \bibfield  {author} {\bibinfo {author} {\bibfnamefont {J.}~\bibnamefont
  {Mecke}}, \bibinfo {author} {\bibfnamefont {J.~O.}\ \bibnamefont {Nketsiah}},
  \bibinfo {author} {\bibfnamefont {R.}~\bibnamefont {Li}},\ and\ \bibinfo
  {author} {\bibfnamefont {Y.}~\bibnamefont {Gao}},\ }\bibfield  {title}
  {\bibinfo {title} {Emergent phenomena in chiral active matter},\ }\href
  {https://doi.org/https://doi.org/10.1360/nso/20230086} {\bibfield  {journal}
  {\bibinfo  {journal} {National Science Open}\ }\textbf {\bibinfo {volume}
  {3}},\ \bibinfo {pages} {20230086} (\bibinfo {year}
  {2024}{\natexlab{b}})}\BibitemShut {NoStop}%
\bibitem [{\citenamefont {Khain}\ \emph {et~al.}(2022)\citenamefont {Khain},
  \citenamefont {Scheibner}, \citenamefont {Fruchart},\ and\ \citenamefont
  {Vitelli}}]{khain2022}%
  \BibitemOpen
  \bibfield  {author} {\bibinfo {author} {\bibfnamefont {T.}~\bibnamefont
  {Khain}}, \bibinfo {author} {\bibfnamefont {C.}~\bibnamefont {Scheibner}},
  \bibinfo {author} {\bibfnamefont {M.}~\bibnamefont {Fruchart}},\ and\
  \bibinfo {author} {\bibfnamefont {V.}~\bibnamefont {Vitelli}},\ }\bibfield
  {title} {\bibinfo {title} {Stokes flows in three-dimensional fluids with odd
  and parity-violating viscosities},\ }\href
  {https://doi.org/10.1017/jfm.2021.1079} {\bibfield  {journal} {\bibinfo
  {journal} {J. Fluid Mech.}\ }\textbf {\bibinfo {volume} {934}},\ \bibinfo
  {pages} {A23} (\bibinfo {year} {2022})}\BibitemShut {NoStop}%
\bibitem [{\citenamefont {Hosaka}\ \emph
  {et~al.}(2023{\natexlab{a}})\citenamefont {Hosaka}, \citenamefont
  {Golestanian},\ and\ \citenamefont {Vilfan}}]{hosaka2023lorentz}%
  \BibitemOpen
  \bibfield  {author} {\bibinfo {author} {\bibfnamefont {Y.}~\bibnamefont
  {Hosaka}}, \bibinfo {author} {\bibfnamefont {R.}~\bibnamefont
  {Golestanian}},\ and\ \bibinfo {author} {\bibfnamefont {A.}~\bibnamefont
  {Vilfan}},\ }\bibfield  {title} {\bibinfo {title} {Lorentz reciprocal theorem
  in fluids with odd viscosity},\ }\href
  {https://doi.org/10.1103/PhysRevLett.131.178303} {\bibfield  {journal}
  {\bibinfo  {journal} {Phys. Rev. Lett.}\ }\textbf {\bibinfo {volume} {131}},\
  \bibinfo {pages} {178303} (\bibinfo {year} {2023}{\natexlab{a}})}\BibitemShut
  {NoStop}%
\bibitem [{\citenamefont {Everts}\ and\ \citenamefont
  {Cichocki}(2024)}]{everts2024dissipative}%
  \BibitemOpen
  \bibfield  {author} {\bibinfo {author} {\bibfnamefont {J.~C.}\ \bibnamefont
  {Everts}}\ and\ \bibinfo {author} {\bibfnamefont {B.}~\bibnamefont
  {Cichocki}},\ }\bibfield  {title} {\bibinfo {title} {Dissipative effects in
  odd viscous {S}tokes flow around a single sphere},\ }\href
  {https://doi.org/https://doi.org/10.1103/PhysRevLett.132.218303} {\bibfield
  {journal} {\bibinfo  {journal} {Phys. Rev. Lett.}\ }\textbf {\bibinfo
  {volume} {132}},\ \bibinfo {pages} {218303} (\bibinfo {year}
  {2024})}\BibitemShut {NoStop}%
\bibitem [{\citenamefont {Mecke}\ \emph {et~al.}(2023)\citenamefont {Mecke},
  \citenamefont {Gao}, \citenamefont {Ram{\'\i}rez~Medina}, \citenamefont
  {Aarts}, \citenamefont {Gompper},\ and\ \citenamefont
  {Ripoll}}]{mecke2023simultaneous}%
  \BibitemOpen
  \bibfield  {author} {\bibinfo {author} {\bibfnamefont {J.}~\bibnamefont
  {Mecke}}, \bibinfo {author} {\bibfnamefont {Y.}~\bibnamefont {Gao}}, \bibinfo
  {author} {\bibfnamefont {C.~A.}\ \bibnamefont {Ram{\'\i}rez~Medina}},
  \bibinfo {author} {\bibfnamefont {D.~G.}\ \bibnamefont {Aarts}}, \bibinfo
  {author} {\bibfnamefont {G.}~\bibnamefont {Gompper}},\ and\ \bibinfo {author}
  {\bibfnamefont {M.}~\bibnamefont {Ripoll}},\ }\bibfield  {title} {\bibinfo
  {title} {Simultaneous emergence of active turbulence and odd viscosity in a
  colloidal chiral active system},\ }\href
  {https://doi.org/10.1038/s42005-023-01442-3} {\bibfield  {journal} {\bibinfo
  {journal} {Commun. Phys.}\ }\textbf {\bibinfo {volume} {6}},\ \bibinfo
  {pages} {324} (\bibinfo {year} {2023})}\BibitemShut {NoStop}%
\bibitem [{\citenamefont {Hargus}\ \emph {et~al.}(2020)\citenamefont {Hargus},
  \citenamefont {Klymko}, \citenamefont {Epstein},\ and\ \citenamefont
  {Mandadapu}}]{hargus2020time}%
  \BibitemOpen
  \bibfield  {author} {\bibinfo {author} {\bibfnamefont {C.}~\bibnamefont
  {Hargus}}, \bibinfo {author} {\bibfnamefont {K.}~\bibnamefont {Klymko}},
  \bibinfo {author} {\bibfnamefont {J.~M.}\ \bibnamefont {Epstein}},\ and\
  \bibinfo {author} {\bibfnamefont {K.~K.}\ \bibnamefont {Mandadapu}},\
  }\bibfield  {title} {\bibinfo {title} {Time reversal symmetry breaking and
  odd viscosity in active fluids: Green--{K}ubo and {NEMD} results},\ }\href
  {https://doi.org/10.1063/5.0006441} {\bibfield  {journal} {\bibinfo
  {journal} {J. Chem. Phys.}\ }\textbf {\bibinfo {volume} {152}},\ \bibinfo
  {pages} {201102} (\bibinfo {year} {2020})}\BibitemShut {NoStop}%
\bibitem [{\citenamefont {Han}\ \emph {et~al.}(2021)\citenamefont {Han},
  \citenamefont {Fruchart}, \citenamefont {Scheibner}, \citenamefont
  {Vaikuntanathan}, \citenamefont {de~Pablo},\ and\ \citenamefont
  {Vitelli}}]{han2021fluctuating}%
  \BibitemOpen
  \bibfield  {author} {\bibinfo {author} {\bibfnamefont {M.}~\bibnamefont
  {Han}}, \bibinfo {author} {\bibfnamefont {M.}~\bibnamefont {Fruchart}},
  \bibinfo {author} {\bibfnamefont {C.}~\bibnamefont {Scheibner}}, \bibinfo
  {author} {\bibfnamefont {S.}~\bibnamefont {Vaikuntanathan}}, \bibinfo
  {author} {\bibfnamefont {J.~J.}\ \bibnamefont {de~Pablo}},\ and\ \bibinfo
  {author} {\bibfnamefont {V.}~\bibnamefont {Vitelli}},\ }\bibfield  {title}
  {\bibinfo {title} {Fluctuating hydrodynamics of chiral active fluids},\
  }\href {https://doi.org/10.1038/s41567-021-01360-7} {\bibfield  {journal}
  {\bibinfo  {journal} {Nat. Phys.}\ }\textbf {\bibinfo {volume} {17}},\
  \bibinfo {pages} {1260} (\bibinfo {year} {2021})}\BibitemShut {NoStop}%
\bibitem [{\citenamefont {Lou}\ \emph {et~al.}(2022)\citenamefont {Lou},
  \citenamefont {Yang}, \citenamefont {Ding}, \citenamefont {Liu},
  \citenamefont {Chen}, \citenamefont {Zhou}, \citenamefont {Ye}, \citenamefont
  {Podgornik},\ and\ \citenamefont {Yang}}]{lou2022odd}%
  \BibitemOpen
  \bibfield  {author} {\bibinfo {author} {\bibfnamefont {X.}~\bibnamefont
  {Lou}}, \bibinfo {author} {\bibfnamefont {Q.}~\bibnamefont {Yang}}, \bibinfo
  {author} {\bibfnamefont {Y.}~\bibnamefont {Ding}}, \bibinfo {author}
  {\bibfnamefont {P.}~\bibnamefont {Liu}}, \bibinfo {author} {\bibfnamefont
  {K.}~\bibnamefont {Chen}}, \bibinfo {author} {\bibfnamefont {X.}~\bibnamefont
  {Zhou}}, \bibinfo {author} {\bibfnamefont {F.}~\bibnamefont {Ye}}, \bibinfo
  {author} {\bibfnamefont {R.}~\bibnamefont {Podgornik}},\ and\ \bibinfo
  {author} {\bibfnamefont {M.}~\bibnamefont {Yang}},\ }\bibfield  {title}
  {\bibinfo {title} {Odd viscosity-induced {H}all-like transport of an active
  chiral fluid},\ }\href {https://doi.org/10.1073/pnas.2201279119} {\bibfield
  {journal} {\bibinfo  {journal} {Proc. Natl. Acad. Sci. U.S.A.}\ }\textbf
  {\bibinfo {volume} {119}},\ \bibinfo {pages} {e2201279119} (\bibinfo {year}
  {2022})}\BibitemShut {NoStop}%
\bibitem [{\citenamefont {Caporusso}\ \emph {et~al.}(2024)\citenamefont
  {Caporusso}, \citenamefont {Gonnella},\ and\ \citenamefont
  {Levis}}]{caporusso2024phase}%
  \BibitemOpen
  \bibfield  {author} {\bibinfo {author} {\bibfnamefont {C.~B.}\ \bibnamefont
  {Caporusso}}, \bibinfo {author} {\bibfnamefont {G.}~\bibnamefont
  {Gonnella}},\ and\ \bibinfo {author} {\bibfnamefont {D.}~\bibnamefont
  {Levis}},\ }\bibfield  {title} {\bibinfo {title} {{Phase coexistence and edge
  currents in the chiral Lennard-Jones fluid}},\ }\href
  {https://doi.org/https://doi.org/10.1103/PhysRevLett.132.168201} {\bibfield
  {journal} {\bibinfo  {journal} {Phys. Rev. Lett.}\ }\textbf {\bibinfo
  {volume} {132}},\ \bibinfo {pages} {168201} (\bibinfo {year}
  {2024})}\BibitemShut {NoStop}%
\bibitem [{\citenamefont {Ding}\ \emph {et~al.}(2024)\citenamefont {Ding},
  \citenamefont {Wang}, \citenamefont {Yang}, \citenamefont {Zhao},
  \citenamefont {Komura}, \citenamefont {Seto}, \citenamefont {Yang},\ and\
  \citenamefont {Ye}}]{ding2024odd}%
  \BibitemOpen
  \bibfield  {author} {\bibinfo {author} {\bibfnamefont {Y.}~\bibnamefont
  {Ding}}, \bibinfo {author} {\bibfnamefont {B.}~\bibnamefont {Wang}}, \bibinfo
  {author} {\bibfnamefont {Q.}~\bibnamefont {Yang}}, \bibinfo {author}
  {\bibfnamefont {Z.}~\bibnamefont {Zhao}}, \bibinfo {author} {\bibfnamefont
  {S.}~\bibnamefont {Komura}}, \bibinfo {author} {\bibfnamefont
  {R.}~\bibnamefont {Seto}}, \bibinfo {author} {\bibfnamefont {M.}~\bibnamefont
  {Yang}},\ and\ \bibinfo {author} {\bibfnamefont {F.}~\bibnamefont {Ye}},\
  }\bibfield  {title} {\bibinfo {title} {Odd response-induced phase separation
  of active spinners},\ }\href {https://doi.org/10.34133/research.0356}
  {\bibfield  {journal} {\bibinfo  {journal} {Research}\ }\textbf {\bibinfo
  {volume} {7}},\ \bibinfo {pages} {0356} (\bibinfo {year} {2024})}\BibitemShut
  {NoStop}%
\bibitem [{\citenamefont {Hargus}\ \emph {et~al.}(2021)\citenamefont {Hargus},
  \citenamefont {Epstein},\ and\ \citenamefont {Mandadapu}}]{hargus2021odd}%
  \BibitemOpen
  \bibfield  {author} {\bibinfo {author} {\bibfnamefont {C.}~\bibnamefont
  {Hargus}}, \bibinfo {author} {\bibfnamefont {J.~M.}\ \bibnamefont
  {Epstein}},\ and\ \bibinfo {author} {\bibfnamefont {K.~K.}\ \bibnamefont
  {Mandadapu}},\ }\bibfield  {title} {\bibinfo {title} {Odd diffusivity of
  chiral random motion},\ }\href
  {https://doi.org/https://doi.org/10.1103/PhysRevLett.127.178001} {\bibfield
  {journal} {\bibinfo  {journal} {Phys. Rev. Lett.}\ }\textbf {\bibinfo
  {volume} {127}},\ \bibinfo {pages} {178001} (\bibinfo {year}
  {2021})}\BibitemShut {NoStop}%
\bibitem [{\citenamefont {Kalz}\ \emph {et~al.}(2022)\citenamefont {Kalz},
  \citenamefont {Vuijk}, \citenamefont {Abdoli}, \citenamefont {Sommer},
  \citenamefont {L{\"o}wen},\ and\ \citenamefont
  {Sharma}}]{kalz2022collisions}%
  \BibitemOpen
  \bibfield  {author} {\bibinfo {author} {\bibfnamefont {E.}~\bibnamefont
  {Kalz}}, \bibinfo {author} {\bibfnamefont {H.~D.}\ \bibnamefont {Vuijk}},
  \bibinfo {author} {\bibfnamefont {I.}~\bibnamefont {Abdoli}}, \bibinfo
  {author} {\bibfnamefont {J.-U.}\ \bibnamefont {Sommer}}, \bibinfo {author}
  {\bibfnamefont {H.}~\bibnamefont {L{\"o}wen}},\ and\ \bibinfo {author}
  {\bibfnamefont {A.}~\bibnamefont {Sharma}},\ }\bibfield  {title} {\bibinfo
  {title} {Collisions enhance self-diffusion in odd-diffusive systems},\ }\href
  {https://doi.org/https://doi.org/10.1103/PhysRevLett.129.090601} {\bibfield
  {journal} {\bibinfo  {journal} {Phys. Rev. Lett.}\ }\textbf {\bibinfo
  {volume} {129}},\ \bibinfo {pages} {090601} (\bibinfo {year}
  {2022})}\BibitemShut {NoStop}%
\bibitem [{\citenamefont {Scheibner}\ \emph {et~al.}(2020)\citenamefont
  {Scheibner}, \citenamefont {Souslov}, \citenamefont {Banerjee}, \citenamefont
  {Sur{\'o}wka}, \citenamefont {Irvine},\ and\ \citenamefont
  {Vitelli}}]{scheibner2020odd}%
  \BibitemOpen
  \bibfield  {author} {\bibinfo {author} {\bibfnamefont {C.}~\bibnamefont
  {Scheibner}}, \bibinfo {author} {\bibfnamefont {A.}~\bibnamefont {Souslov}},
  \bibinfo {author} {\bibfnamefont {D.}~\bibnamefont {Banerjee}}, \bibinfo
  {author} {\bibfnamefont {P.}~\bibnamefont {Sur{\'o}wka}}, \bibinfo {author}
  {\bibfnamefont {W.~T.}\ \bibnamefont {Irvine}},\ and\ \bibinfo {author}
  {\bibfnamefont {V.}~\bibnamefont {Vitelli}},\ }\bibfield  {title} {\bibinfo
  {title} {Odd elasticity},\ }\href
  {https://doi.org/https://doi.org/10.1038/s41567-020-0795-y} {\bibfield
  {journal} {\bibinfo  {journal} {Nat. Phys.}\ }\textbf {\bibinfo {volume}
  {16}},\ \bibinfo {pages} {475} (\bibinfo {year} {2020})}\BibitemShut
  {NoStop}%
\bibitem [{\citenamefont {Ganeshan}\ and\ \citenamefont
  {Abanov}(2017)}]{ganeshan2017}%
  \BibitemOpen
  \bibfield  {author} {\bibinfo {author} {\bibfnamefont {S.}~\bibnamefont
  {Ganeshan}}\ and\ \bibinfo {author} {\bibfnamefont {A.~G.}\ \bibnamefont
  {Abanov}},\ }\bibfield  {title} {\bibinfo {title} {Odd viscosity in
  two-dimensional incompressible fluids},\ }\href
  {https://doi.org/10.1103/PhysRevFluids.2.094101} {\bibfield  {journal}
  {\bibinfo  {journal} {Phys. Rev. Fluids}\ }\textbf {\bibinfo {volume} {2}},\
  \bibinfo {pages} {094101} (\bibinfo {year} {2017})}\BibitemShut {NoStop}%
\bibitem [{\citenamefont {Zhao}\ \emph {et~al.}(2021)\citenamefont {Zhao},
  \citenamefont {Wang}, \citenamefont {Komura}, \citenamefont {Yang},
  \citenamefont {Ye},\ and\ \citenamefont {Seto}}]{zhao2021}%
  \BibitemOpen
  \bibfield  {author} {\bibinfo {author} {\bibfnamefont {Z.}~\bibnamefont
  {Zhao}}, \bibinfo {author} {\bibfnamefont {B.}~\bibnamefont {Wang}}, \bibinfo
  {author} {\bibfnamefont {S.}~\bibnamefont {Komura}}, \bibinfo {author}
  {\bibfnamefont {M.}~\bibnamefont {Yang}}, \bibinfo {author} {\bibfnamefont
  {F.}~\bibnamefont {Ye}},\ and\ \bibinfo {author} {\bibfnamefont
  {R.}~\bibnamefont {Seto}},\ }\bibfield  {title} {\bibinfo {title} {Emergent
  stripes of active rotors in shear flows},\ }\href
  {https://doi.org/https://doi.org/10.1103/PhysRevResearch.3.043229} {\bibfield
   {journal} {\bibinfo  {journal} {Phys. Rev. Research}\ }\textbf {\bibinfo
  {volume} {3}},\ \bibinfo {pages} {043229} (\bibinfo {year}
  {2021})}\BibitemShut {NoStop}%
\bibitem [{\citenamefont {Furst}\ and\ \citenamefont
  {Squires}(2017)}]{furst2017microrheology}%
  \BibitemOpen
  \bibfield  {author} {\bibinfo {author} {\bibfnamefont {E.~M.}\ \bibnamefont
  {Furst}}\ and\ \bibinfo {author} {\bibfnamefont {T.~M.}\ \bibnamefont
  {Squires}},\ }\href@noop {} {\emph {\bibinfo {title} {Microrheology}}}\
  (\bibinfo  {publisher} {Oxford University Press},\ \bibinfo {year}
  {2017})\BibitemShut {NoStop}%
\bibitem [{\citenamefont {Zell}\ \emph {et~al.}(2014)\citenamefont {Zell},
  \citenamefont {Nowbahar}, \citenamefont {Mansard}, \citenamefont {Leal},
  \citenamefont {Deshmukh}, \citenamefont {Mecca}, \citenamefont {Tucker},\
  and\ \citenamefont {Squires}}]{zell2014surface}%
  \BibitemOpen
  \bibfield  {author} {\bibinfo {author} {\bibfnamefont {Z.~A.}\ \bibnamefont
  {Zell}}, \bibinfo {author} {\bibfnamefont {A.}~\bibnamefont {Nowbahar}},
  \bibinfo {author} {\bibfnamefont {V.}~\bibnamefont {Mansard}}, \bibinfo
  {author} {\bibfnamefont {L.~G.}\ \bibnamefont {Leal}}, \bibinfo {author}
  {\bibfnamefont {S.~S.}\ \bibnamefont {Deshmukh}}, \bibinfo {author}
  {\bibfnamefont {J.~M.}\ \bibnamefont {Mecca}}, \bibinfo {author}
  {\bibfnamefont {C.~J.}\ \bibnamefont {Tucker}},\ and\ \bibinfo {author}
  {\bibfnamefont {T.~M.}\ \bibnamefont {Squires}},\ }\bibfield  {title}
  {\bibinfo {title} {Surface shear inviscidity of soluble surfactants},\ }\href
  {https://doi.org/https://doi.org/10.1073/pnas.1315991111} {\bibfield
  {journal} {\bibinfo  {journal} {Proc. Natl. Acad. Sci. U.S.A.}\ }\textbf
  {\bibinfo {volume} {111}},\ \bibinfo {pages} {3677} (\bibinfo {year}
  {2014})}\BibitemShut {NoStop}%
\bibitem [{\citenamefont {Barentin}\ \emph {et~al.}(1999)\citenamefont
  {Barentin}, \citenamefont {Ybert}, \citenamefont {Di~Meglio},\ and\
  \citenamefont {Joanny}}]{barentin1999}%
  \BibitemOpen
  \bibfield  {author} {\bibinfo {author} {\bibfnamefont {C.}~\bibnamefont
  {Barentin}}, \bibinfo {author} {\bibfnamefont {C.}~\bibnamefont {Ybert}},
  \bibinfo {author} {\bibfnamefont {J.-M.}\ \bibnamefont {Di~Meglio}},\ and\
  \bibinfo {author} {\bibfnamefont {J.-F.}\ \bibnamefont {Joanny}},\ }\bibfield
   {title} {\bibinfo {title} {Surface shear viscosity of {G}ibbs and {L}angmuir
  monolayers},\ }\href {https://doi.org/10.1017/S0022112099006321} {\bibfield
  {journal} {\bibinfo  {journal} {J. Fluid Mech.}\ }\textbf {\bibinfo {volume}
  {397}},\ \bibinfo {pages} {331} (\bibinfo {year} {1999})}\BibitemShut
  {NoStop}%
\bibitem [{\citenamefont {Crocker}\ \emph {et~al.}(2000)\citenamefont
  {Crocker}, \citenamefont {Valentine}, \citenamefont {Weeks}, \citenamefont
  {Gisler}, \citenamefont {Kaplan}, \citenamefont {Yodh},\ and\ \citenamefont
  {Weitz}}]{crocker2000two}%
  \BibitemOpen
  \bibfield  {author} {\bibinfo {author} {\bibfnamefont {J.~C.}\ \bibnamefont
  {Crocker}}, \bibinfo {author} {\bibfnamefont {M.~T.}\ \bibnamefont
  {Valentine}}, \bibinfo {author} {\bibfnamefont {E.~R.}\ \bibnamefont
  {Weeks}}, \bibinfo {author} {\bibfnamefont {T.}~\bibnamefont {Gisler}},
  \bibinfo {author} {\bibfnamefont {P.~D.}\ \bibnamefont {Kaplan}}, \bibinfo
  {author} {\bibfnamefont {A.~G.}\ \bibnamefont {Yodh}},\ and\ \bibinfo
  {author} {\bibfnamefont {D.~A.}\ \bibnamefont {Weitz}},\ }\bibfield  {title}
  {\bibinfo {title} {Two-point microrheology of inhomogeneous soft materials},\
  }\href {https://doi.org/https://doi.org/10.1103/PhysRevLett.85.888}
  {\bibfield  {journal} {\bibinfo  {journal} {Phys. Rev. Lett.}\ }\textbf
  {\bibinfo {volume} {85}},\ \bibinfo {pages} {888} (\bibinfo {year}
  {2000})}\BibitemShut {NoStop}%
\bibitem [{\citenamefont {Lier}\ \emph {et~al.}(2023)\citenamefont {Lier},
  \citenamefont {Duclut}, \citenamefont {Bo}, \citenamefont {Armas},
  \citenamefont {J{\"u}licher},\ and\ \citenamefont
  {Sur{\'o}wka}}]{lier2023lift}%
  \BibitemOpen
  \bibfield  {author} {\bibinfo {author} {\bibfnamefont {R.}~\bibnamefont
  {Lier}}, \bibinfo {author} {\bibfnamefont {C.}~\bibnamefont {Duclut}},
  \bibinfo {author} {\bibfnamefont {S.}~\bibnamefont {Bo}}, \bibinfo {author}
  {\bibfnamefont {J.}~\bibnamefont {Armas}}, \bibinfo {author} {\bibfnamefont
  {F.}~\bibnamefont {J{\"u}licher}},\ and\ \bibinfo {author} {\bibfnamefont
  {P.}~\bibnamefont {Sur{\'o}wka}},\ }\bibfield  {title} {\bibinfo {title}
  {Lift force in odd compressible fluids},\ }\href
  {https://doi.org/10.1103/PhysRevE.108.L023101} {\bibfield  {journal}
  {\bibinfo  {journal} {Phys. Rev. E}\ }\textbf {\bibinfo {volume} {108}},\
  \bibinfo {pages} {L023101} (\bibinfo {year} {2023})}\BibitemShut {NoStop}%
\bibitem [{\citenamefont {Duclut}\ \emph {et~al.}(2024)\citenamefont {Duclut},
  \citenamefont {Bo}, \citenamefont {Lier}, \citenamefont {Armas},
  \citenamefont {Sur{\'o}wka},\ and\ \citenamefont
  {J{\"u}licher}}]{duclut2024probe}%
  \BibitemOpen
  \bibfield  {author} {\bibinfo {author} {\bibfnamefont {C.}~\bibnamefont
  {Duclut}}, \bibinfo {author} {\bibfnamefont {S.}~\bibnamefont {Bo}}, \bibinfo
  {author} {\bibfnamefont {R.}~\bibnamefont {Lier}}, \bibinfo {author}
  {\bibfnamefont {J.}~\bibnamefont {Armas}}, \bibinfo {author} {\bibfnamefont
  {P.}~\bibnamefont {Sur{\'o}wka}},\ and\ \bibinfo {author} {\bibfnamefont
  {F.}~\bibnamefont {J{\"u}licher}},\ }\bibfield  {title} {\bibinfo {title}
  {Probe particles in odd active viscoelastic fluids: How activity and
  dissipation determine linear stability},\ }\href
  {https://doi.org/https://doi.org/10.1103/PhysRevE.109.044126} {\bibfield
  {journal} {\bibinfo  {journal} {Phys. Rev. E}\ }\textbf {\bibinfo {volume}
  {109}},\ \bibinfo {pages} {044126} (\bibinfo {year} {2024})}\BibitemShut
  {NoStop}%
\bibitem [{\citenamefont {Epstein}\ and\ \citenamefont
  {Mandadapu}(2020)}]{epstein2020}%
  \BibitemOpen
  \bibfield  {author} {\bibinfo {author} {\bibfnamefont {J.~M.}\ \bibnamefont
  {Epstein}}\ and\ \bibinfo {author} {\bibfnamefont {K.~K.}\ \bibnamefont
  {Mandadapu}},\ }\bibfield  {title} {\bibinfo {title} {Time-reversal symmetry
  breaking in two-dimensional nonequilibrium viscous fluids},\ }\href
  {https://doi.org/https://doi.org/10.1103/PhysRevE.101.052614} {\bibfield
  {journal} {\bibinfo  {journal} {Phys. Rev. E}\ }\textbf {\bibinfo {volume}
  {101}},\ \bibinfo {pages} {052614} (\bibinfo {year} {2020})}\BibitemShut
  {NoStop}%
\bibitem [{\citenamefont {Markovich}\ \emph {et~al.}(2019)\citenamefont
  {Markovich}, \citenamefont {Tjhung},\ and\ \citenamefont
  {Cates}}]{markovich2019chiral}%
  \BibitemOpen
  \bibfield  {author} {\bibinfo {author} {\bibfnamefont {T.}~\bibnamefont
  {Markovich}}, \bibinfo {author} {\bibfnamefont {E.}~\bibnamefont {Tjhung}},\
  and\ \bibinfo {author} {\bibfnamefont {M.~E.}\ \bibnamefont {Cates}},\
  }\bibfield  {title} {\bibinfo {title} {Chiral active matter: microscopic
  'torque dipoles' have more than one hydrodynamic description},\ }\href
  {https://doi.org/10.1088/1367-2630/ab54af} {\bibfield  {journal} {\bibinfo
  {journal} {New J. Phys.}\ }\textbf {\bibinfo {volume} {21}},\ \bibinfo
  {pages} {112001} (\bibinfo {year} {2019})}\BibitemShut {NoStop}%
\bibitem [{\citenamefont {Elfring}\ \emph {et~al.}(2016)\citenamefont
  {Elfring}, \citenamefont {Leal},\ and\ \citenamefont
  {Squires}}]{elfring2016surface}%
  \BibitemOpen
  \bibfield  {author} {\bibinfo {author} {\bibfnamefont {G.~J.}\ \bibnamefont
  {Elfring}}, \bibinfo {author} {\bibfnamefont {L.~G.}\ \bibnamefont {Leal}},\
  and\ \bibinfo {author} {\bibfnamefont {T.~M.}\ \bibnamefont {Squires}},\
  }\bibfield  {title} {\bibinfo {title} {Surface viscosity and {Marangoni}
  stresses at surfactant laden interfaces},\ }\href
  {https://doi.org/https://doi.org/10.1017/jfm.2016.96} {\bibfield  {journal}
  {\bibinfo  {journal} {J. Fluid Mech.}\ }\textbf {\bibinfo {volume} {792}},\
  \bibinfo {pages} {712} (\bibinfo {year} {2016})}\BibitemShut {NoStop}%
\bibitem [{\citenamefont {Manikantan}\ and\ \citenamefont
  {Squires}(2020)}]{manikantan2020surfactant}%
  \BibitemOpen
  \bibfield  {author} {\bibinfo {author} {\bibfnamefont {H.}~\bibnamefont
  {Manikantan}}\ and\ \bibinfo {author} {\bibfnamefont {T.~M.}\ \bibnamefont
  {Squires}},\ }\bibfield  {title} {\bibinfo {title} {Surfactant dynamics:
  hidden variables controlling fluid flows},\ }\href
  {https://doi.org/10.1017/jfm.2020.170} {\bibfield  {journal} {\bibinfo
  {journal} {J. Fluid Mech.}\ }\textbf {\bibinfo {volume} {892}},\ \bibinfo
  {pages} {P1} (\bibinfo {year} {2020})}\BibitemShut {NoStop}%
\bibitem [{\citenamefont {Hosaka}\ \emph {et~al.}(2021)\citenamefont {Hosaka},
  \citenamefont {Komura},\ and\ \citenamefont
  {Andelman}}]{hosaka2021nonreciprocal}%
  \BibitemOpen
  \bibfield  {author} {\bibinfo {author} {\bibfnamefont {Y.}~\bibnamefont
  {Hosaka}}, \bibinfo {author} {\bibfnamefont {S.}~\bibnamefont {Komura}},\
  and\ \bibinfo {author} {\bibfnamefont {D.}~\bibnamefont {Andelman}},\
  }\bibfield  {title} {\bibinfo {title} {Nonreciprocal response of a
  two-dimensional fluid with odd viscosity},\ }\href
  {https://doi.org/10.1103/PhysRevE.103.042610} {\bibfield  {journal} {\bibinfo
   {journal} {Phys. Rev. E}\ }\textbf {\bibinfo {volume} {103}},\ \bibinfo
  {pages} {042610} (\bibinfo {year} {2021})}\BibitemShut {NoStop}%
\bibitem [{\citenamefont {Hosaka}\ \emph
  {et~al.}(2023{\natexlab{b}})\citenamefont {Hosaka}, \citenamefont
  {Golestanian},\ and\ \citenamefont
  {Daddi-Moussa-Ider}}]{hosaka2023hydrodynamics}%
  \BibitemOpen
  \bibfield  {author} {\bibinfo {author} {\bibfnamefont {Y.}~\bibnamefont
  {Hosaka}}, \bibinfo {author} {\bibfnamefont {R.}~\bibnamefont
  {Golestanian}},\ and\ \bibinfo {author} {\bibfnamefont {A.}~\bibnamefont
  {Daddi-Moussa-Ider}},\ }\bibfield  {title} {\bibinfo {title} {Hydrodynamics
  of an odd active surfer in a chiral fluid},\ }\href
  {https://doi.org/10.1088/1367-2630/aceea4} {\bibfield  {journal} {\bibinfo
  {journal} {New J. Phys.}\ }\textbf {\bibinfo {volume} {25}},\ \bibinfo
  {pages} {083046} (\bibinfo {year} {2023}{\natexlab{b}})}\BibitemShut
  {NoStop}%
\bibitem [{\citenamefont {Daddi-Moussa-Ider}\ \emph {et~al.}(2025)\citenamefont
  {Daddi-Moussa-Ider}, \citenamefont {Vilfan},\ and\ \citenamefont
  {Hosaka}}]{daddi2025analytical}%
  \BibitemOpen
  \bibfield  {author} {\bibinfo {author} {\bibfnamefont {A.}~\bibnamefont
  {Daddi-Moussa-Ider}}, \bibinfo {author} {\bibfnamefont {A.}~\bibnamefont
  {Vilfan}},\ and\ \bibinfo {author} {\bibfnamefont {Y.}~\bibnamefont
  {Hosaka}},\ }\bibfield  {title} {\bibinfo {title} {Analytical solution for
  the hydrodynamic resistance of a disk in a compressible fluid layer with odd
  viscosity on a rigid substrate},\ }\href {https://doi.org/10.1063/5.0249623}
  {\bibfield  {journal} {\bibinfo  {journal} {J. Chem. Phys.}\ }\textbf
  {\bibinfo {volume} {162}},\ \bibinfo {pages} {064103} (\bibinfo {year}
  {2025})}\BibitemShut {NoStop}%
\bibitem [{\citenamefont {Daddi-Moussa-Ider}\ and\ \citenamefont
  {Gekle}(2016)}]{daddi2016hydrodynamic}%
  \BibitemOpen
  \bibfield  {author} {\bibinfo {author} {\bibfnamefont {A.}~\bibnamefont
  {Daddi-Moussa-Ider}}\ and\ \bibinfo {author} {\bibfnamefont {S.}~\bibnamefont
  {Gekle}},\ }\bibfield  {title} {\bibinfo {title} {Hydrodynamic interaction
  between particles near elastic interfaces},\ }\href
  {https://doi.org/10.1063/1.4955099} {\bibfield  {journal} {\bibinfo
  {journal} {J. Chem. Phys.}\ }\textbf {\bibinfo {volume} {145}},\ \bibinfo
  {pages} {014905} (\bibinfo {year} {2016})}\BibitemShut {NoStop}%
\bibitem [{\citenamefont {Felderhof}(2006)}]{felderhof2006dynamics}%
  \BibitemOpen
  \bibfield  {author} {\bibinfo {author} {\bibfnamefont {B.}~\bibnamefont
  {Felderhof}},\ }\bibfield  {title} {\bibinfo {title} {Dynamics of an
  interface with adsorption layer between two fluids},\ }\href
  {https://doi.org/10.1063/1.2372460} {\bibfield  {journal} {\bibinfo
  {journal} {Phys. Fluids}\ }\textbf {\bibinfo {volume} {18}},\ \bibinfo
  {pages} {112103} (\bibinfo {year} {2006})}\BibitemShut {NoStop}%
\bibitem [{\citenamefont {Daddi-Moussa-Ider}\ and\ \citenamefont
  {Gekle}(2018)}]{daddi2018brownian}%
  \BibitemOpen
  \bibfield  {author} {\bibinfo {author} {\bibfnamefont {A.}~\bibnamefont
  {Daddi-Moussa-Ider}}\ and\ \bibinfo {author} {\bibfnamefont {S.}~\bibnamefont
  {Gekle}},\ }\bibfield  {title} {\bibinfo {title} {Brownian motion near an
  elastic cell membrane: A theoretical study},\ }\href
  {https://doi.org/https://doi.org/10.1140/epje/i2018-11627-6} {\bibfield
  {journal} {\bibinfo  {journal} {Eur. Phys. J. E}\ }\textbf {\bibinfo {volume}
  {41}},\ \bibinfo {pages} {19} (\bibinfo {year} {2018})}\BibitemShut {NoStop}%
\bibitem [{\citenamefont {Daddi-Moussa-Ider}\ \emph {et~al.}(2018)\citenamefont
  {Daddi-Moussa-Ider}, \citenamefont {Lisicki}, \citenamefont {Gekle},
  \citenamefont {Menzel},\ and\ \citenamefont
  {L{\"o}wen}}]{daddi2018hydrodynamic}%
  \BibitemOpen
  \bibfield  {author} {\bibinfo {author} {\bibfnamefont {A.}~\bibnamefont
  {Daddi-Moussa-Ider}}, \bibinfo {author} {\bibfnamefont {M.}~\bibnamefont
  {Lisicki}}, \bibinfo {author} {\bibfnamefont {S.}~\bibnamefont {Gekle}},
  \bibinfo {author} {\bibfnamefont {A.~M.}\ \bibnamefont {Menzel}},\ and\
  \bibinfo {author} {\bibfnamefont {H.}~\bibnamefont {L{\"o}wen}},\ }\bibfield
  {title} {\bibinfo {title} {Hydrodynamic coupling and rotational mobilities
  near planar elastic membranes},\ }\href {https://doi.org/10.1063/1.5032304}
  {\bibfield  {journal} {\bibinfo  {journal} {J. Chem. Phys.}\ }\textbf
  {\bibinfo {volume} {149}},\ \bibinfo {pages} {014901} (\bibinfo {year}
  {2018})}\BibitemShut {NoStop}%
\bibitem [{\citenamefont {Daddi-Moussa-Ider}\ \emph {et~al.}(2019)\citenamefont
  {Daddi-Moussa-Ider}, \citenamefont {Kurzthaler}, \citenamefont {Hoell},
  \citenamefont {Z{\"o}ttl}, \citenamefont {Mirzakhanloo}, \citenamefont
  {Alam}, \citenamefont {Menzel}, \citenamefont {L{\"o}wen},\ and\
  \citenamefont {Gekle}}]{daddi2019frequency}%
  \BibitemOpen
  \bibfield  {author} {\bibinfo {author} {\bibfnamefont {A.}~\bibnamefont
  {Daddi-Moussa-Ider}}, \bibinfo {author} {\bibfnamefont {C.}~\bibnamefont
  {Kurzthaler}}, \bibinfo {author} {\bibfnamefont {C.}~\bibnamefont {Hoell}},
  \bibinfo {author} {\bibfnamefont {A.}~\bibnamefont {Z{\"o}ttl}}, \bibinfo
  {author} {\bibfnamefont {M.}~\bibnamefont {Mirzakhanloo}}, \bibinfo {author}
  {\bibfnamefont {M.-R.}\ \bibnamefont {Alam}}, \bibinfo {author}
  {\bibfnamefont {A.~M.}\ \bibnamefont {Menzel}}, \bibinfo {author}
  {\bibfnamefont {H.}~\bibnamefont {L{\"o}wen}},\ and\ \bibinfo {author}
  {\bibfnamefont {S.}~\bibnamefont {Gekle}},\ }\bibfield  {title} {\bibinfo
  {title} {Frequency-dependent higher-order {Stokes} singularities near a
  planar elastic boundary: Implications for the hydrodynamics of an active
  microswimmer near an elastic interface},\ }\href@noop {} {\bibfield
  {journal} {\bibinfo  {journal} {Phys. Rev. E}\ }\textbf {\bibinfo {volume}
  {100}},\ \bibinfo {pages} {032610} (\bibinfo {year} {2019})}\BibitemShut
  {NoStop}%
\bibitem [{\citenamefont {Daddi-Moussa-Ider}\ \emph
  {et~al.}(2024{\natexlab{a}})\citenamefont {Daddi-Moussa-Ider}, \citenamefont
  {Tjhung}, \citenamefont {Richter},\ and\ \citenamefont
  {Menzel}}]{Daddi-Moussa-Ider_2024_JPCM}%
  \BibitemOpen
  \bibfield  {author} {\bibinfo {author} {\bibfnamefont {A.}~\bibnamefont
  {Daddi-Moussa-Ider}}, \bibinfo {author} {\bibfnamefont {E.}~\bibnamefont
  {Tjhung}}, \bibinfo {author} {\bibfnamefont {T.}~\bibnamefont {Richter}},\
  and\ \bibinfo {author} {\bibfnamefont {A.~M.}\ \bibnamefont {Menzel}},\
  }\bibfield  {title} {\bibinfo {title} {Hydrodynamics of a disk in a thin film
  of weakly nematic fluid subject to linear friction},\ }\href
  {https://doi.org/10.1088/1361-648X/ad65ad} {\bibfield  {journal} {\bibinfo
  {journal} {J. Phys.: Condens. Matter}\ }\textbf {\bibinfo {volume} {36}},\
  \bibinfo {pages} {445101} (\bibinfo {year} {2024}{\natexlab{a}})}\BibitemShut
  {NoStop}%
\bibitem [{\citenamefont {Daddi-Moussa-Ider}\ \emph
  {et~al.}(2024{\natexlab{b}})\citenamefont {Daddi-Moussa-Ider}, \citenamefont
  {Tjhung}, \citenamefont {Pradas}, \citenamefont {Richter},\ and\
  \citenamefont {Menzel}}]{daddi2024rotational}%
  \BibitemOpen
  \bibfield  {author} {\bibinfo {author} {\bibfnamefont {A.}~\bibnamefont
  {Daddi-Moussa-Ider}}, \bibinfo {author} {\bibfnamefont {E.}~\bibnamefont
  {Tjhung}}, \bibinfo {author} {\bibfnamefont {M.}~\bibnamefont {Pradas}},
  \bibinfo {author} {\bibfnamefont {T.}~\bibnamefont {Richter}},\ and\ \bibinfo
  {author} {\bibfnamefont {A.~M.}\ \bibnamefont {Menzel}},\ }\bibfield  {title}
  {\bibinfo {title} {Rotational dynamics of a disk in a thin film of weakly
  nematic fluid subject to linear friction},\ }\href
  {https://doi.org/https://doi.org/10.1140/epje/s10189-024-00452-5} {\bibfield
  {journal} {\bibinfo  {journal} {Eur. Phys. J. E}\ }\textbf {\bibinfo {volume}
  {47}},\ \bibinfo {pages} {58} (\bibinfo {year}
  {2024}{\natexlab{b}})}\BibitemShut {NoStop}%
\bibitem [{\citenamefont {Daddi-Moussa-Ider}\ \emph
  {et~al.}(2024{\natexlab{c}})\citenamefont {Daddi-Moussa-Ider}, \citenamefont
  {Golestanian},\ and\ \citenamefont {Vilfan}}]{daddi2024hydrodynamic_JFM}%
  \BibitemOpen
  \bibfield  {author} {\bibinfo {author} {\bibfnamefont {A.}~\bibnamefont
  {Daddi-Moussa-Ider}}, \bibinfo {author} {\bibfnamefont {R.}~\bibnamefont
  {Golestanian}},\ and\ \bibinfo {author} {\bibfnamefont {A.}~\bibnamefont
  {Vilfan}},\ }\bibfield  {title} {\bibinfo {title} {Hydrodynamic efficiency
  limit on a {Marangoni} surfer},\ }\href
  {https://doi.org/https://doi.org/10.1017/jfm.2024.363} {\bibfield  {journal}
  {\bibinfo  {journal} {J. Fluid Mech.}\ }\textbf {\bibinfo {volume} {986}},\
  \bibinfo {pages} {A32} (\bibinfo {year} {2024}{\natexlab{c}})}\BibitemShut
  {NoStop}%
\bibitem [{\citenamefont {Abramowitz}\ and\ \citenamefont
  {Stegun}(2000)}]{abramowitz2000handbook}%
  \BibitemOpen
  \bibfield  {author} {\bibinfo {author} {\bibfnamefont {M.}~\bibnamefont
  {Abramowitz}}\ and\ \bibinfo {author} {\bibfnamefont {I.~A.}\ \bibnamefont
  {Stegun}},\ }\href@noop {} {\emph {\bibinfo {title} {{Handbook of
  Mathematical Functions with Formulas, Graphs, and Mathematical Tables}}}},\
  Vol.~\bibinfo {volume} {55}\ (\bibinfo  {publisher} {Dover Publications Inc,
  New York},\ \bibinfo {year} {2000})\BibitemShut {NoStop}%
\bibitem [{\citenamefont {Baddour}(2011)}]{baddour2011two}%
  \BibitemOpen
  \bibfield  {author} {\bibinfo {author} {\bibfnamefont {N.}~\bibnamefont
  {Baddour}},\ }\bibfield  {title} {\bibinfo {title} {Two-dimensional {Fourier}
  transforms in polar coordinates},\ }in\ \href
  {https://doi.org/https://doi.org/10.1016/B978-0-12-385861-0.00001-4} {\emph
  {\bibinfo {booktitle} {Adv. Imaging Electron Phys.}}},\ Vol.\ \bibinfo
  {volume} {165}\ (\bibinfo  {publisher} {Elsevier},\ \bibinfo {year} {2011})\
  pp.\ \bibinfo {pages} {1--45}\BibitemShut {NoStop}%
\bibitem [{\citenamefont {Spiegel}\ \emph {et~al.}(2009)\citenamefont
  {Spiegel}, \citenamefont {Lipschutz}, \citenamefont {Schiller},\ and\
  \citenamefont {Spellman}}]{spiegel2009complex}%
  \BibitemOpen
  \bibfield  {author} {\bibinfo {author} {\bibfnamefont {M.~R.}\ \bibnamefont
  {Spiegel}}, \bibinfo {author} {\bibfnamefont {S.}~\bibnamefont {Lipschutz}},
  \bibinfo {author} {\bibfnamefont {J.~J.}\ \bibnamefont {Schiller}},\ and\
  \bibinfo {author} {\bibfnamefont {D.}~\bibnamefont {Spellman}},\ }\href@noop
  {} {\emph {\bibinfo {title} {{Complex Variables}}}}\ (\bibinfo  {publisher}
  {McGraw Hill, New York},\ \bibinfo {year} {2009})\BibitemShut {NoStop}%
\bibitem [{\citenamefont {Happel}\ and\ \citenamefont
  {Brenner}(1983)}]{happel2012low}%
  \BibitemOpen
  \bibfield  {author} {\bibinfo {author} {\bibfnamefont {J.}~\bibnamefont
  {Happel}}\ and\ \bibinfo {author} {\bibfnamefont {H.}~\bibnamefont
  {Brenner}},\ }\href {https://doi.org/10.1007/978-94-009-8352-6} {\emph
  {\bibinfo {title} {Low {R}eynolds Number Hydrodynamics}}}\ (\bibinfo
  {publisher} {Springer},\ \bibinfo {address} {Netherlands},\ \bibinfo {year}
  {1983})\BibitemShut {NoStop}%
\bibitem [{\citenamefont {Chwang}\ and\ \citenamefont
  {Wu}(1974)}]{chwang_wu_1974}%
  \BibitemOpen
  \bibfield  {author} {\bibinfo {author} {\bibfnamefont {A.~T.}\ \bibnamefont
  {Chwang}}\ and\ \bibinfo {author} {\bibfnamefont {T.~Y.-T.}\ \bibnamefont
  {Wu}},\ }\bibfield  {title} {\bibinfo {title} {Hydromechanics of
  low-{R}eynolds-number flow. {Part 1}. {R}otation of axisymmetric prolate
  bodies},\ }\href {https://doi.org/10.1017/S0022112074001819} {\bibfield
  {journal} {\bibinfo  {journal} {J. Fluid Mech.}\ }\textbf {\bibinfo {volume}
  {63}},\ \bibinfo {pages} {607} (\bibinfo {year} {1974})}\BibitemShut
  {NoStop}%
\bibitem [{\citenamefont {Hosaka}\ \emph {et~al.}(2024)\citenamefont {Hosaka},
  \citenamefont {Chatzittofi}, \citenamefont {Golestanian},\ and\ \citenamefont
  {Vilfan}}]{hosaka2024chirotactic}%
  \BibitemOpen
  \bibfield  {author} {\bibinfo {author} {\bibfnamefont {Y.}~\bibnamefont
  {Hosaka}}, \bibinfo {author} {\bibfnamefont {M.}~\bibnamefont {Chatzittofi}},
  \bibinfo {author} {\bibfnamefont {R.}~\bibnamefont {Golestanian}},\ and\
  \bibinfo {author} {\bibfnamefont {A.}~\bibnamefont {Vilfan}},\ }\bibfield
  {title} {\bibinfo {title} {Chirotactic response of microswimmers in fluids
  with odd viscosity},\ }\href
  {https://doi.org/https://doi.org/10.1103/PhysRevResearch.6.L032044}
  {\bibfield  {journal} {\bibinfo  {journal} {Phys. Rev. Research}\ }\textbf
  {\bibinfo {volume} {6}},\ \bibinfo {pages} {L032044} (\bibinfo {year}
  {2024})}\BibitemShut {NoStop}%
\bibitem [{\citenamefont {Khain}\ \emph {et~al.}(2024)\citenamefont {Khain},
  \citenamefont {Fruchart}, \citenamefont {Scheibner}, \citenamefont {Witten},\
  and\ \citenamefont {Vitelli}}]{khain2024trading}%
  \BibitemOpen
  \bibfield  {author} {\bibinfo {author} {\bibfnamefont {T.}~\bibnamefont
  {Khain}}, \bibinfo {author} {\bibfnamefont {M.}~\bibnamefont {Fruchart}},
  \bibinfo {author} {\bibfnamefont {C.}~\bibnamefont {Scheibner}}, \bibinfo
  {author} {\bibfnamefont {T.~A.}\ \bibnamefont {Witten}},\ and\ \bibinfo
  {author} {\bibfnamefont {V.}~\bibnamefont {Vitelli}},\ }\bibfield  {title}
  {\bibinfo {title} {Trading particle shape with fluid symmetry: on the
  mobility matrix in 3-{D} chiral fluids},\ }\href
  {https://doi.org/https://doi.org/10.1017/jfm.2024.535} {\bibfield  {journal}
  {\bibinfo  {journal} {J. Fluid Mech.}\ }\textbf {\bibinfo {volume} {992}},\
  \bibinfo {pages} {A5} (\bibinfo {year} {2024})}\BibitemShut {NoStop}%
\bibitem [{\citenamefont {Evans}\ and\ \citenamefont
  {Sackmann}(1988)}]{evans1988}%
  \BibitemOpen
  \bibfield  {author} {\bibinfo {author} {\bibfnamefont {E.}~\bibnamefont
  {Evans}}\ and\ \bibinfo {author} {\bibfnamefont {E.}~\bibnamefont
  {Sackmann}},\ }\bibfield  {title} {\bibinfo {title} {Translational and
  rotational drag coefficients for a disk moving in a liquid membrane
  associated with a rigid substrate},\ }\href
  {https://doi.org/10.1017/S0022112088003106} {\bibfield  {journal} {\bibinfo
  {journal} {J. Fluid Mech.}\ }\textbf {\bibinfo {volume} {194}},\ \bibinfo
  {pages} {553} (\bibinfo {year} {1988})}\BibitemShut {NoStop}%
\bibitem [{\citenamefont {Guazzelli}\ and\ \citenamefont
  {Morris}(2011)}]{guazzelli2011physical}%
  \BibitemOpen
  \bibfield  {author} {\bibinfo {author} {\bibfnamefont {E.}~\bibnamefont
  {Guazzelli}}\ and\ \bibinfo {author} {\bibfnamefont {J.~F.}\ \bibnamefont
  {Morris}},\ }\href {https://doi.org/https://doi.org/10.1017/CBO9780511894671}
  {\emph {\bibinfo {title} {{A Physical Introduction to Suspension
  Dynamics}}}}\ (\bibinfo  {publisher} {Cambridge University Press},\ \bibinfo
  {year} {2011})\BibitemShut {NoStop}%
\bibitem [{\citenamefont {Lapa}\ and\ \citenamefont {Hughes}(2014)}]{lapa2014}%
  \BibitemOpen
  \bibfield  {author} {\bibinfo {author} {\bibfnamefont {M.~F.}\ \bibnamefont
  {Lapa}}\ and\ \bibinfo {author} {\bibfnamefont {T.~L.}\ \bibnamefont
  {Hughes}},\ }\bibfield  {title} {\bibinfo {title} {Swimming at low {Reynolds}
  number in fluids with odd, or {Hall}, viscosity},\ }\href
  {https://doi.org/10.1103/PhysRevE.89.043019} {\bibfield  {journal} {\bibinfo
  {journal} {Phys. Rev. E}\ }\textbf {\bibinfo {volume} {89}},\ \bibinfo
  {pages} {043019} (\bibinfo {year} {2014})}\BibitemShut {NoStop}%
\bibitem [{\citenamefont {Brown}\ and\ \citenamefont
  {Churchill}(2009)}]{brown2009complex}%
  \BibitemOpen
  \bibfield  {author} {\bibinfo {author} {\bibfnamefont {J.~W.}\ \bibnamefont
  {Brown}}\ and\ \bibinfo {author} {\bibfnamefont {R.~V.}\ \bibnamefont
  {Churchill}},\ }\href@noop {} {\emph {\bibinfo {title} {{Complex Variables
  and Applications}}}}\ (\bibinfo  {publisher} {McGraw-Hill, New York},\
  \bibinfo {year} {2009})\BibitemShut {NoStop}%
\end{thebibliography}
%

\end{document}